\documentclass[apj,numberedappendix]{emulateapj}
\usepackage{graphicx}
\usepackage{amsmath}

\def\solmass{$M_{\sun}$}

\def\solperyr{$M_{\sun}$ yr$^{-1}$}
\def\solperpc{$M_{\sun}$ pc$^{-2}$}

\newcommand{\halpha}{H$\alpha$}

\newcommand{\HI}{H{\sc i}}
\newcommand{\Htwo}{H$_2$}
\newcommand{\SigHI}{$\Sigma_{\mathrm{atomic}}$}
\newcommand{\SigHtwo}{$\Sigma_{\mathrm{mol}}$}
\newcommand{\SigGas}{$\Sigma_{\mathrm{gas}}$}
\newcommand{\SigSFR}{$\Sigma_{\mathrm{SFR}}$}
\newcommand{\SigStar}{$\Sigma_{*}$}
\newcommand{\um}{$\mu$m}
\newcommand{\tdep}{$\tau_{\rm dep}$}
\newcommand{\Rmol}{$R_{\rm mol}$}

\newcommand{\Pmp}{$P_{\rm mp}$}
\begin{document}

\title{Panchromatic Hubble Andromeda Treasury XVI. Star Cluster Formation Efficiency and the Clustered Fraction of Young Stars}

\author{L. Clifton Johnson\altaffilmark{1,2}, Anil C. Seth\altaffilmark{3}, Julianne J. Dalcanton\altaffilmark{2}, Lori C. Beerman\altaffilmark{2}, Morgan Fouesneau\altaffilmark{4}, Alexia R. Lewis\altaffilmark{2}, Daniel R. Weisz\altaffilmark{2,9}, Benjamin F. Williams\altaffilmark{2}, Eric F. Bell\altaffilmark{5}, Andrew E. Dolphin\altaffilmark{6}, S{\o}ren S. Larsen\altaffilmark{7}, Karin Sandstrom\altaffilmark{1}, Evan D. Skillman\altaffilmark{8}}

\email{lcj@ucsd.edu}
\affil{$^{1}$Center for Astrophysics and Space Sciences, University of California, San Diego, 9500 Gilman Drive, La Jolla, CA 92093, USA}
\affil{$^{2}$Department of Astronomy, University of Washington, Box 351580, Seattle, WA 98195, USA}
\affil{$^{3}$Department of Physics and Astronomy, University of Utah, Salt Lake City, UT 84112, USA}
\affil{$^{4}$Max-Planck-Institut f\"ur Astronomie, K\"onigstuhl 17, 69117 Heidelberg, Germany}
\affil{$^{5}$Department of Astronomy, University of Michigan, 1085 South University Avenue, Ann Arbor, MI 48109, USA}
\affil{$^{6}$Raytheon Company, 1151 East Hermans Road, Tucson, AZ 85756, USA}
\affil{$^{7}$Department of Astrophysics, IMAPP, Radboud University Nijmegen, P.O. Box 9010, 6500 GL Nijmegen, The Netherlands}
\affil{$^{8}$Minnesota Institute for Astrophysics, University of Minnesota, 116 Church Street SE, Minneapolis, MN 55455, USA}
\altaffiltext{9}{Hubble Fellow}


\begin{abstract}
We use the Panchromatic Hubble Andromeda Treasury (PHAT) survey dataset to perform spatially resolved measurements of star cluster formation efficiency ($\Gamma$), the fraction of stellar mass formed in long-lived star clusters. We use robust star formation history and cluster parameter constraints, obtained through color-magnitude diagram analysis of resolved stellar populations, to study Andromeda's cluster and field populations over the last $\sim$300 Myr.  We measure $\Gamma$ of 4--8\% for young, 10--100 Myr old populations in M31.  We find that cluster formation efficiency varies systematically across the M31 disk, consistent with variations in mid-plane pressure.  These $\Gamma$ measurements expand the range of well-studied galactic environments, providing precise constraints in an \HI-dominated, low intensity star formation environment.  Spatially resolved results from M31 are broadly consistent with previous trends observed on galaxy-integrated scales, where $\Gamma$ increases with increasing star formation rate surface density (\SigSFR).  However, we can explain observed scatter in the relation and attain better agreement between observations and theoretical models if we account for environmental variations in gas depletion time (\tdep) when modeling $\Gamma$, accounting for the qualitative shift in star formation behavior when transitioning from a \Htwo-dominated to a \HI-dominated interstellar medium.  We also demonstrate that $\Gamma$ measurements in high \SigSFR\ starburst systems are well-explained by \tdep-dependent fiducial $\Gamma$ models.
\end{abstract}


\section{Introduction} \label{intro}

The clustering behavior of stars is a direct, observable result of star formation physics.  At the onset of star formation, young embedded stars inherit the highly structured spatial distribution of the molecular gas from which they form.  The newly formed stars soon decouple from the gas due to stellar feedback processes.  Because star formation is an inefficient process \citep[$\sim$1\% per free-fall time;][]{Krumholz12}, gas dispersal removes most of a region's binding gravitational potential.  This results in the distribution of stars expanding and dispersing, creating stellar associations and complexes with characteristic sizes of tens to hundreds of parsecs.  In some cases, however, the concentration of stellar mass is high enough that collections of stars remain gravitationally bound and tightly clustered beyond the initial gas embedded phase, creating long-lived ($\gtrsim$10 Myr) star clusters that we observe today.

Observations of star clusters provide the means to constrain theoretical descriptions of star formation.  Star cluster formation depends on the complex interplay of: 1) star formation efficiency, which dictates how much of the gas reservoir is transformed into possible cluster members; 2) stellar feedback processes, which drive the transition from gas-rich to gas-poor local environments; and 3) the energetics of the natal environment, which determine the kinematics of stellar and gaseous components within the star forming region.  As a result, accurately reproducing the observed behavior of star clusters, and young stellar distributions generally, is a key challenge for any theoretical star formation model.

We explore an important observational metric of star cluster formation in this work: star cluster formation efficiency, which is the fraction of stellar mass born in long-lived star clusters \citep[$\Gamma$;][]{Bastian08, AdamoBastian15}.  This quantity directly relates cluster formation to total star formation activity.  Past measurements of cluster formation efficiency have been obtained on galaxy-integrated scales for a wide range of galaxies \citep[e.g., ][]{Larsen00, Goddard10, SilvaVilla11, Adamo11, Cook12}.  These studies provided evidence that $\Gamma$ varies systematically as a function of star forming environment, quantified according to star formation rate surface density, \SigSFR.  Cluster formation efficiencies range from a few percent for galaxies with low star formation intensity up to $\sim$50\% for high intensity galaxy mergers.  Recently, studies have begun to explore $\Gamma$ with increasing detail, performing spatially resolved analyses to better investigate the environmental dependence of cluster formation \citep{SilvaVilla13, Ryon14, Adamo15}.

In addition to these observational studies, work from \citet{Kruijssen12} took an important first theoretical step in modeling and predicting the behavior of $\Gamma$.  Building on the theoretical work of \citet{Elmegreen08} and star formation simulations by \citet{Bonnell08}, \citet{Kruijssen12} presents a framework to predict $\Gamma$ based on the idea that long-lived star clusters emerge from regions with high star formation efficiency. In this model, the densest portions of hierarchically-structured molecular clouds attain high star formation efficiencies because while the star formation efficiency remains constant per free-fall time \citep{Elmegreen02}, these regions progress through multiple short free-fall times.  As a result, these regions become stellar-dominated before gas expulsion truncates star formation.  Low gas fractions in these dense sub-regions prevent subsequent gas expulsion from dramatically changing the gravitational potential, leading to the formation of long-lived star clusters.

In this work, we measure star cluster formation efficiency across the Andromeda galaxy (M31) using data from the Hubble Space Telescope (HST) obtained by the Panchromatic Hubble Andromeda Treasury survey \citep[PHAT;][]{Dalcanton12}.  M31 is an interesting target of investigation for a number of reasons.  First, Andromeda hosts a relatively low intensity star formation environment, characterized by small values of \SigSFR.  The galaxy's modest level of star formation bolsters the range of parameter space where $\Gamma$ has been measured, providing good contrast with active star forming galaxies previously studied \citep[e.g., M83;][]{Adamo15}.  Second, M31's predominantly atomic phase interstellar medium (ISM) sets it apart from most previous $\Gamma$ analysis targets, which are typically dominated by their molecular phase.  Finally, preliminary investigations show that Andromeda's cluster dissolution rate is low \citep{Fouesneau14}, suggesting characteristic disruption timescales $>$100--300 Myr that leave its population of long-lived star clusters intact for study.

Our analysis of M31 benefits from a number of important advantages over previous extragalactic $\Gamma$ studies.  First, we use a robust catalog of 2753 clusters that were visually identified as part of the Andromeda Project citizen science project \citep{Johnson15_AP}.  This cluster search was performed on uniform imaging from the PHAT survey, in which clusters appear as groupings of individually resolved member stars, reducing confusion and ambiguity in cluster identifications with respect to ground-based surveys of M31 or HST-based surveys of more distant galaxies.  In imaging of galaxies at larger distances ($>$1 Mpc), cluster members are blended together and cluster profiles are only marginally resolved, even with the resolving power of HST.  Second, catalog completeness is well characterized and shows that the PHAT young cluster sample ($<$300 Myr old) is complete to 500--1000 \solmass\ (depending on age and galactic position), providing access to low mass clusters that are undetectable in most extragalactic surveys.  Finally, the ability to resolve individual cluster member stars permits the use of color-magnitude diagram (CMD) fitting to derive cluster ages and masses.  This fitting technique provides stronger constraints than those obtained through multi-band integrated light SED fitting of young clusters, particularly for low mass clusters where large stochastic variations in the integrated light are common \citep[see e.g.,][]{Fouesneau10, Krumholz15}.

The benefits of studying cluster formation efficiency in M31 reach beyond the realm of cluster-specific observations.  Star formation history (SFH) results derived from the PHAT observations of field star populations provide valuable spatially ($\sim$100 pc scales) and temporally ($\Delta$ log Age/yr $\sim$ 0.1) resolved information about the total star formation activity across the disk of M31 \citep{Lewis15}. These constraints are a considerable improvement over emission line and multiwavelength total star formation rate (SFR) estimates (e.g., via \halpha, FUV+24$\mu$m).  In addition, the availability of \HI\ and CO datasets with high spatial resolution and sensitivity allow the detailed characterization of the star forming ISM, even at low gas surface densities.  These gas phase constraints provide rich ancillary information that allow us to map how differences in natal environments affect properties of emergent cluster populations.

In this paper, we take advantage of the superior quality of data provided by the PHAT survey to perform a high precision, spatially resolved investigation of $\Gamma$ across a range of star forming environments in M31.  Our work complements previous observational studies by providing a high quality anchor to extragalactic $\Gamma$ measurements in more distant galaxies where the level of detail available with respect to characterizing both clusters and field populations is limited by available spatial resolution.

This paper is organized into six sections.  We begin with a description of the observational data in Section \ref{data}, followed by a presentation of star cluster and field star characterization analysis in Sections \ref{analysis_cluster} and \ref{analysis_sfh}.  We calculate $\Gamma$ and its associated uncertainties in Section \ref{results_gamma}.  In Section \ref{results_theory}, we calculate theoretical predictions for $\Gamma$ using the \citet{Kruijssen12} model and compare these to the observational results.  In Section \ref{discuss}, we compare our results to previous observations, discuss the validity of key assumptions, and follow-up on interesting aspects of our results and their broader implications.  We finish with a summary of our work in Section \ref{summary}.  Throughout this study, we assume a distance modulus for M31 of 24.47 \citep[785 kpc;][]{McConnachie05}, for which 1~arcsec corresponds to a physical size of 3.81~pc.  

\subsection{$\Gamma$ and Cluster Definitions} \label{definitions}

We begin our study by clearly defining the measurement we pursue in this work.  The quantity of interest here is the fraction of stellar mass born in long-lived star clusters relative to the total coeval stellar mass formed.  We refer to this value as $\Gamma$ following \citet{Bastian08}, and this ratio is equivalently referred to as the cluster formation efficiency. We rewrite the original definition presented by Bastian et~al.\ (i.e., the ratio of the cluster formation rate to the total star formation rate, CFR/SFR) to clarify that this measurement is made over a specific age interval.  We define:
\begin{equation} \label{gamdef}
\Gamma = \frac{M_{\rm cl,tot} (t_1, t_2)}{M_{\rm tot} (t_1, t_2)},
\end{equation}
where $M_{\rm cl,tot}$ represents the integrated cluster mass, $M_{\rm tot}$ represents the integrated total stellar mass, and ($t_1$, $t_2$) represents the age interval over which the masses are integrated.

The focus on long-lived star clusters is specified in order to distinguish this set of gravitationally bound stellar systems (a class that includes open clusters, globular clusters, and young massive clusters) from two other distinct types of objects: embedded clusters and stellar associations.  \citet{Lada03} found that while $\sim$90\% of stars are formed in embedded clusters in the Solar neighborhood, only a small fraction ($\Gamma \sim$ 4--7\%) of these systems survive gas expulsion and become long-lived gravitationally bound star clusters.  The remaining unbound stars disperse and go on to form stellar associations and large scale star forming complexes.

Catalog contamination from embedded clusters is naturally avoided in optical wavelength cluster searches due to the fact that their large attenuations make these objects undetectable.  On the other hand, differentiating between gravitationally bound star clusters and unbound associations is often a difficult task, and one that is critical to the accurate assessment of cluster formation efficiency. Differences in adopted cluster definitions and sample selection has previously lead to conflicting results in the literature concerning cluster age distributions and dissolution timescales \citep[e.g.,][]{Chandar10-M83, Bastian12}, as well as $\Gamma$ values \citep{Chandar15, Kruijssen16}.  Following an approach similar to the one recommended in \citet{Kruijssen16}, we reduce or eliminate contamination from unbound associations through a careful choice of the analyzed age interval.

We could adopt a young age limit ($t_1$) as young as 1--3 Myr, when clusters transition out of their embedded phase, distinguish themselves as long-lived stellar systems, and become detectable in optically-selected samples.  In practice, however, it is difficult to differentiate between long-lived, gravitationally bound star clusters and unbound, expanding stellar associations that are still compact at young ages.  It is difficult to differentiate between the two types of objects using only spatial distributions and stellar surface densities until the stars have time to dynamically evolve.  

Fortunately, associations expand on short timescales.  A study by \citet{Gieles11} demonstrated that by 10 Myr, the distinction between clusters and associations is clear.  By this age, the ratio of a cluster's age to its crossing time ($\Pi \equiv$ Age / $T_{\rm cross}$) increases to values $>$1, while associations continually expand leading to $\Pi \le 1$.  In agreement with this result, we show in Section \ref{clusterfitting} that nearly all PHAT cluster identifications with ages $>$10 Myr have $\Pi$ values $>$1.  We adopt $t_1$=10 Myr for our study to avoid subjective classifications of ambiguous young stellar systems.  From a practical standpoint, this choice has few downsides.  Excluding cluster and field populations with ages $<$10 Myr only excludes a small fraction of the total $<$300 Myr stellar population that is available for study in the PHAT dataset, while reducing contamination to negligible levels.

In terms of upper age limits, investigators typically restrict measurements of $\Gamma$ to young ages ($t_2$ $<$ 10--100 Myr) for a number of reasons.  First, SFR estimates obtained from broadband indicators (e.g., \halpha, FUV+24$\mu$m) or from fitting shallow color-magnitude diagrams only provide constraints at young ages.  Second, estimates of total cluster mass are increasingly reliant on mass function extrapolations and small number statistics with increasing age due to evolutionary fading and rising mass completeness limits.  Finally, dynamical mass loss and cluster dissolution are smallest at young ages, while at older ages observations of $\Gamma$ may no longer reflect its initial value.  With the PHAT data, we can measure $\Gamma$ out to older ages ($t_2$ = 300 Myr) thanks to deep optical imaging that allows detection of main sequence (MS) turnoffs in both clusters and the field down to $\sim$3 \solmass.

We choose to measure $\Gamma$ over two age ranges: 10--100 Myr and 100--300 Myr.  The minimum and maximum values $t_1$=10 Myr and $t_2$=300 Myr are set by the limitations of the PHAT dataset as discussed above.  We primarily focus on 10--100 Myr $\Gamma$ measurements throughout this study due to better available time resolution for the SFH, better age precision for the clusters, as well as the compatibility of this age range with previous studies.  In addition, measurements in this younger age bin should correspond better to present-day ISM properties and \citet{Kruijssen12} predictions.  However, we also analyze the older 100--300 Myr age bin to search for any indication that $\Gamma$ evolves with time.

For the $\Gamma$ measurements in this work, we assume there has been no cluster dissolution over the relevant 10--300 Myr age range.  Analysis of the PHAT cluster age distribution \citep[A. Seth et~al., in preparation;][]{Fouesneau14} appears consistent with little or no cluster destruction within the young cluster population.  Under this assumption, the value of $\Gamma$ should not change with time, and therefore:
\begin{equation}
\Gamma = \Gamma_0 = \Gamma_{10-100} = \Gamma_{100-300}
\end{equation}
where $\Gamma_0$ represents an initial, intrinsic cluster formation efficiency, and $\Gamma_{10-100}$ and $\Gamma_{100-300}$ represent clustered stellar fractions over age ranges of 10--100 Myr and 100--300 Myr.  We assess the validity of this assumption and how cluster dissolution would affect our inference of $\Gamma_0$ in Section \ref{discuss_dissolution}.

\section{Data} \label{data}

\subsection{PHAT Observations and Photometry} \label{data_phat}

The PHAT survey imaged 1/3 of the disk of M31 in six passbands spanning near-ultraviolet to near-infrared wavelengths.  The survey provides resolved stellar photometry of 117 million sources that we use to determine the properties of both the cluster and field populations, with completeness limits that allow the detection of individual MS stars down to $\sim$3 \solmass.  Here we provide an overview of the crowded field stellar photometry derived for PHAT; full details are found in \citet{Dalcanton12} and \citet{Williams14}.

All PHAT resolved stellar photometry is derived using the DOLPHOT software package, an updated version of HSTPHOT \citep{Dolphin00}.  In this work, we use only the optical wavelength Advanced Camera for Surveys (ACS) data, obtained in the F475W and F814W passbands (similar to $g$ and $I$, respectively).

To fit the SFH for the field populations, \citet{Lewis15} used two-band optical photometry catalogs from first PHAT generation survey photometry.  These \texttt{gst} catalogs include high-quality detections that pass criteria for signal-to-noise ratio (S/N), crowding, and sharpness, using photometry parameters described in \citet{Dalcanton12}.  

We characterized clusters using photometry catalogs that are tailored to highly crowded cluster environments, and differ from the field star catalogs in two ways.  First, the two-band optical photometry was calculated using the revised photometry parameters described in \citet{Williams14}.  Second, we adopted a set of quality cuts for the cluster catalogs that are less strict than those used for the field \texttt{gst} catalogs: S/N $>$ 4 in both passbands, (\texttt{Sharp}$_{\rm F475W}$+\texttt{Sharp}$_{\rm F814W}$)$^2$ $\le$ 0.1, and no crowding cut.  

The photometry catalogs are supplemented by large numbers of artificial star tests (ASTs).  The AST results are used to quantify photometric biases, uncertainties, and completeness across the CMD.  The cluster and field ASTs are described in further detail in Sections \ref{clusterfitting} and \ref{analysis_sfh}.

\subsection{Spatial Analysis Regions} \label{data_reg}

To measure $\Gamma$ and investigate its variation across the disk of M31, we divide the PHAT survey footprint into seven regions.  We define these regions according to three considerations: the cumulative mass of young stars required to make a statistically significant measurement of $\Gamma$ due to stochastic sampling of the cluster mass function ($\gtrsim$10$^6$ \solmass); the physical scales associated with young stellar complexes; and the variation of galactic environments in M31.

\begin{figure}
\centering
\includegraphics[scale=0.45]{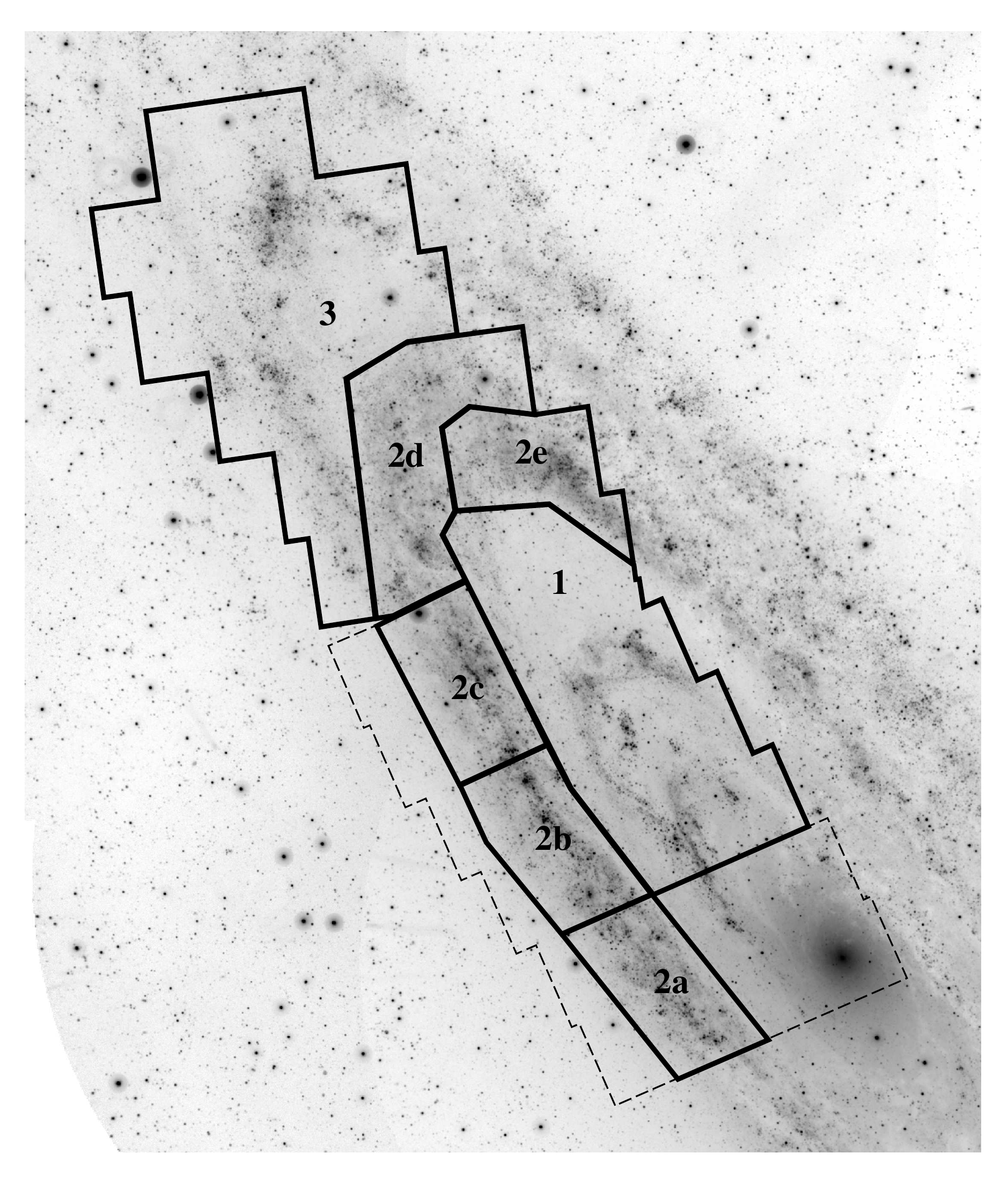}
\caption[M31 Spatial Analysis Regions]{Spatial distribution of M31 analysis regions with labels. The underlying GALEX NUV image highlights young star forming regions.  North is up and east is left in the image.  The dashed line represents the PHAT survey footprint.  Outer disk regions east of the 10 kpc ring and inner disk regions surrounding the galaxy nucleus are excluded from analysis (see Section \ref{data_reg}).}
\label{fig_reg}
\end{figure}

The locations of the seven analysis regions we adopt are shown in Figure \ref{fig_reg}.  These regions were defined to isolate the 10 kpc star forming ring (Region 2) from the inner disk (Region 1; $R_{\rm gc}<$10 kpc) and outer disk (Region 3; $R_{\rm gc}>$13 kpc), and divide the mass formed over the 10--100 Myr age range into approximately equal amounts.  As a result, each region hosts $\sim$3--5 $\times$ 10$^6$ \solmass\ of star formation during the 10--100 Myr epoch.  In addition to isolating the inner and outer disk environments, we further divide the 10 kpc star forming ring into five parts, and isolate two prolific star forming regions: OB54 (Region 2e) and OB30/31 (Region 2a), as identified by \citet{vdB64}.  In addition to the seven primary analysis regions, we also derive results for the 10 kpc ring as a whole, and report survey-wide results by integrating over all seven analysis regions.

We omit a central bulge-dominated region in the inner disk of M31 from our analysis due to increased levels of crowding that degrade the effective depth of the data and make SFH derivations more uncertain \citep{Lewis15}.  We also exclude the outer disk region beyond the eastern portion of the star forming ring because we cannot cleanly separate the ring and outer disk components due to projection effects.  Excluding these regions does not impact our results due to the negligible number of young clusters and total recent star formation that we omit.

\subsection{Ancillary Data and ISM Properties} \label{data_radio}

In addition to the PHAT survey data, we make use of \HI\ observations from WSRT/GBT \citep{Braun09} and $^{12}$CO(1-0) observations from IRAM \citep{Nieten06} to assess properties of the ISM within M31.  The \HI\ and CO datasets have native angular resolution of 30 arcsec and 23 arcsec, respectively.  We refer the reader to the primary references for a full description of the observations and data reduction.

We derive basic properties of M31's ISM within each of the spatial analysis regions using these \HI\ and CO datasets.  For the \HI\ data, we convert column density maps derived by \citet{Braun09} directly to deprojected atomic gas surface density (\SigHI) assuming an inclination angle of 77 degrees and a factor of 1.36 correction to account for helium mass.  We measure molecular gas surface densities (\SigHtwo) using the CO map, making the same inclination correction and adopting a CO-to-H$_2$ conversion consistent with observational constraints from the Milky Way \citep{Bolatto13}: $\alpha_{\rm CO}=4.35$ \solmass\ pc$^{-2}$ (K km s$^{-1}$)$^{-1}$, which assumes $X_{\rm CO}=2 \times 10^{20}$ cm$^{-2}$ (K km s$^{-1}$)$^{-1}$ and already includes a correction for helium.  Next, we smooth the \SigHI\ and \SigHtwo\ maps using a deprojected 0.5 kpc$^2$ measurement kernel (an ellipse with major and minor axes of $\sim$100 and 23 arcsec, respectively).  This smoothing provides symmetric measurements in the deprojected spatial plane, facilitates comparisons to other extragalactic studies that probe kpc-scale surface densities, and allows for a common spatial resolution for analysis of gas and star formation surface densities (see Section \ref{sigsfr}).  Finally, we combine the \SigHI\ and \SigHtwo\ maps to calculate total gas surface densities, \SigGas=\SigHI+\SigHtwo.

We use the newly derived atomic, molecular, and total gas surface density maps to calculate characteristic \SigHI, \SigHtwo, and \SigGas\ values for each of the analysis regions.  We compute mass-weighted average surface densities for each region, where we weight each line of sight by its integrated gas mass.  Using a weighted average, the characteristic surface density values we derive are minimally affected by non-uniform spatial distributions of gas within a region and the specific boundaries used to define the analysis regions.  We discuss weighted surface density measurements in regards to \SigSFR\ calculations in Section \ref{sigsfr}, and expand discussion on this issue in Appendix \ref{appendix_sigma}.  

In agreement with previous work, we find that the ISM throughout M31 is dominated by its atomic component.  We find that the \Htwo-to-\HI\ ratio, $R_{\rm mol} \equiv$ \SigHtwo/\SigHI, ranges from 0.02--0.60, and measurements of \SigGas\ vary between 5--12 \solperpc.  We report region-by-region ISM measurements in Table \ref{tbl_ismobs}.

The IRAM CO data coverage does not extend beyond the star forming ring, therefore we supplement our knowledge of molecular gas in the outer disk region of M31 using high resolution (5 arcsec) interferometric observations of $^{12}$CO(1-0) from CARMA (A. Schruba et~al., in preparation) obtained for a 300 arcsec diameter region in the vicinity of the OB102 star forming complex.  We use these additional observations to estimate \SigHtwo\ and $R_{\rm mol}$ in the outer disk (Region 3).  We obtain \SigHtwo\ of 0.8 \solperpc\ and $R_{\rm mol}$ of 2\%, but acknowledge that these values likely underestimate the surface density of molecular gas due to the lack of short spacing CO observations for this region.  We will regard the measured values of \SigHtwo\ and $R_{\rm mol}$ for the outer disk as lower limits.  Fortunately, the molecular fraction for the outer disk is small ($<$10\%), and therefore the predominantly \HI-based \SigGas\ measurement provides an accurate estimate for the region.

In addition to gas surface densities, we measure $\sigma_{\rm gas}$ values using maps of \HI\ non-thermal velocity dispersion from \citet{Braun09}.  We find little spatial variation in mass-weighted $\sigma_{\rm gas}$ measurements, spanning a range of 7--10 km s$^{-1}$.  We report these measurements in Table \ref{tbl_ismobs}.

\section{Analysis} \label{analysis}

In this section, we describe how we compute $\Gamma$ and its constituent parts, $M_{\rm cl,tot}$ and $M_{\rm tot}$.  We discuss CMD fitting used to measure cluster ages and masses in Section \ref{analysis_cluster}, discuss CMD fitting used to measure total SFHs in Section \ref{analysis_sfh}, and outline our probabilistic $\Gamma$ analysis techniques in Section \ref{analysis_gamma}.

\subsection{Cluster Properties} \label{analysis_cluster}

\subsubsection{PHAT Clusters: Catalog and Completeness} \label{clustercat}

We analyze a cluster sample derived from the Andromeda Project (AP) cluster catalog \citep{Johnson15_AP}.  This catalog includes 2753 star clusters that lie within the PHAT survey footprint, covering a wide range of ages and masses.  These clusters were identified through visual inspection of optical (F475W, F814W) images by volunteer citizen scientists, facilitated through the Zooniverse's Andromeda Project website\footnote{\url{http://www.andromedaproject.org}}.  Each image was examined $>$80 times, providing robust classification statistics for each cluster candidate.  The final sample of clusters was selected according to the fraction of user-weighted cluster identifications using a catalog threshold that maximizes completeness and minimizes contamination with respect to the expert-derived PHAT Year 1 cluster catalog \citep{Johnson12}.

\begin{figure*}
\centering
\includegraphics[scale=0.55]{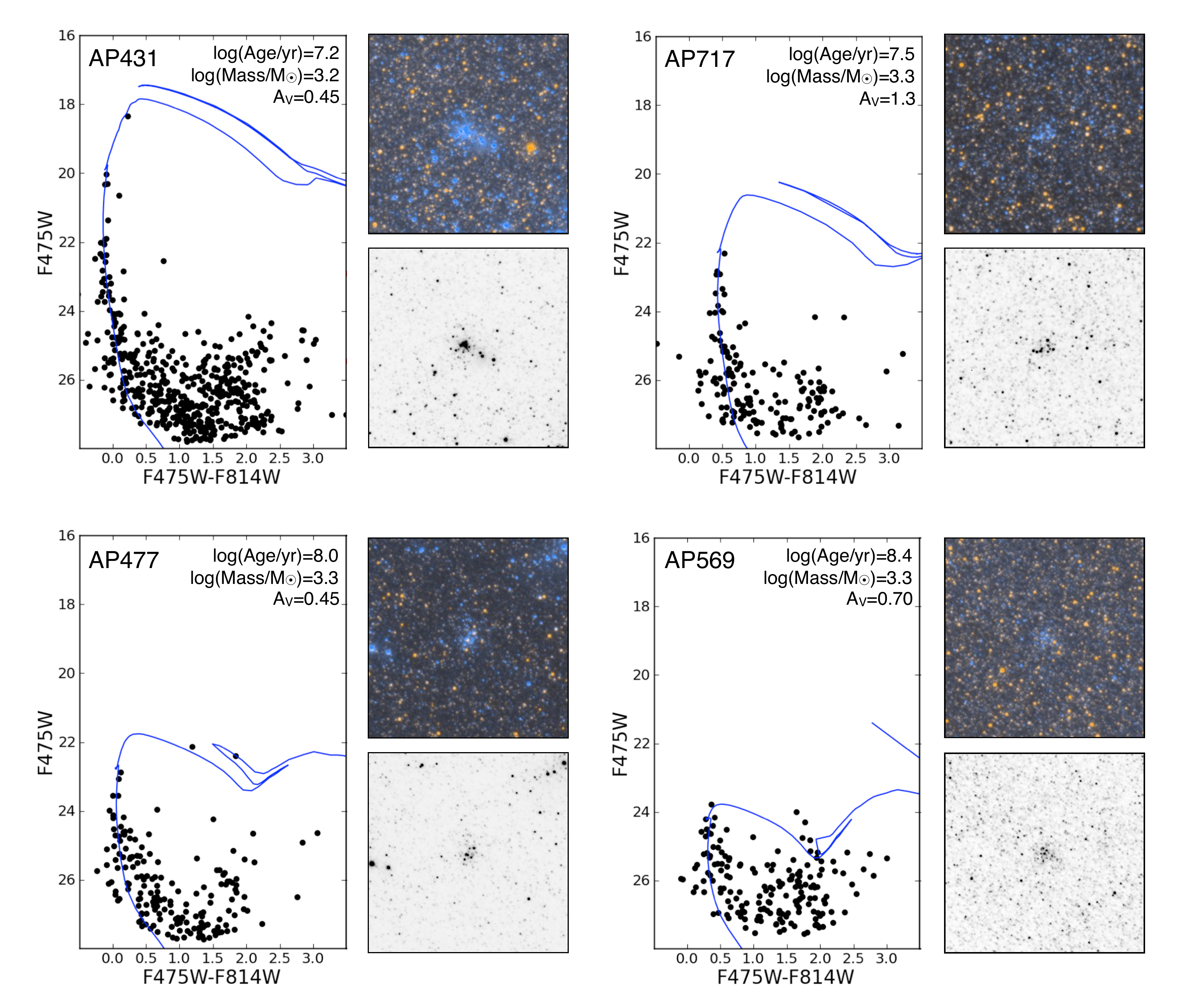}
\caption[Example PHAT Cluster CMDs]{CMDs and cutout images of four example clusters sampling the age range of interest for our $\Gamma$ analysis.  These clusters were chosen for their similar masses of $\sim$2$\times$10$^3$ \solmass\ and their logarithmic spacing in age between 10--300 Myr; fitted parameters for each cluster are listed in the figure.  CMDs include all stars that lie within the cluster's photometric aperture: cluster members and background field stars. Isochrones from the Padova group \citep{Marigo08,Girardi10} representing the best fit age and $A_V$ from MATCH are overlaid on the cluster CMDs.  The color cutout is a F475W+F814W composite, the B/W cutout is an inverted version of a F475W image, and both are 15 arcsec ($\sim$60 pc) on a side.}
\label{fig_exclst}
\end{figure*}

Young star clusters appear in PHAT imaging as collections of individually resolved member stars, as seen for four example clusters presented in Figure \ref{fig_exclst}.  For ages $<$300--500 Myr, the stellar MS is readily detectable, providing robust age constraints for young clusters.  At older ages, red clump and red giant branch (RGB) member stars are still individually resolved, but these features provide limited age information and lead to uninformative constraints.

A critical component of $\Gamma$ analysis is the extrapolation from the observed cluster mass to the total mass of the cluster population.  We use a suite of 3000 synthetic cluster tests to compute catalog completeness and accurately estimate the contribution of undetected low mass clusters to the total cluster mass.  Synthetic clusters were injected into AP search images and passed through the same cluster identification processing as all the real data.  The sample of synthetic clusters covers a wide range of cluster properties (age, mass, dust attenuation) and are distributed throughout the survey footprint to assess cluster detection across a variety of galactic environments.  For a detailed description of the completeness test sample, please see Section 2.2 of \citet{Johnson15_AP}.

We calculate completeness functions for each analysis region in terms of cluster mass, averaged over the two age ranges of interest: 10--100 and 100--300 Myr.  For each region, we select a subsample of synthetic clusters whose input ages and local RGB stellar densities fall within each of the two age bins and the observed range of background stellar densities found within the analysis region.  This selection accounts for the fact that cluster detection not only depends on cluster mass, but also on age and local stellar density of the underlying background.  Due to the structure of M31's stellar disk, the RGB stellar density selection is roughly equivalent to one based on galactocentric radius.  Next, the selected synthetic clusters are assigned weights according to their local MS stellar densities.  This weighting helps account for the difference in spatial distribution between uniformly distributed synthetic cluster tests and the clumpy distribution of young clusters that are biased towards regions of greater stellar density, and hence lower levels of completeness.  Once the synthetic sample is selected and weighted, we model the completeness function in terms of individual cluster mass ($m$) using a logistic function, parameterized by a 50\% completeness limit ($m_{\rm lim}$) and slope parameter ($s_{\rm lim}$):
\begin{equation}
f(m) = [ 1 + \exp(-s_{\rm lim} \times \log_{10}(m/m_{\rm lim}) ) ]^{-1}.
\end{equation}
We fit the synthetic results and report the best fit 50\% completeness limit and slope parameter for each analysis region in Table \ref{tbl_obs}.  These limits range from 520--950 \solmass\ for the 10--100 Myr age bin and from 650--1250 \solmass\ for the 100--300 Myr age bin, depending on position within M31.

\subsubsection{Determining Cluster Ages and Masses} \label{clusterfitting}

CMD fitting of individually resolved member stars provides valuable constraints on a cluster's age and mass.  We use the MATCH software package to analyze cluster CMDs following techniques described in \citet{Dolphin02}.  This software models the observed CMD by simulating stellar populations convolved with observed photometric noise, bias, and completeness.  The code populates theoretical isochrones according to input parameters that define the age, total mass, and dust attenuation of the population, as well as its distance, metallicity, stellar IMF, and binary fraction.  We fit the cluster CMDs assuming a simple stellar population (SSP) model, a special case of SFH fitting where only single-age populations (not linear combinations of multiple populations) are considered.  Synthetic populations are created from unique combinations of age and other input parameters, which are then convolved with a model of observational errors derived from ASTs and combined with a background model (here, representing non-cluster field populations) to produce a simulated CMD distribution.  This simulated CMD is scaled according to total stellar mass (or equivalently, the SFR of the single age bin) and compared to the observed CMD, where the fit quality is evaluated according to a Poisson likelihood function.  The software iterates through a series of synthetic CMDs to estimate the relative likelihood of different combinations of input parameters.

For cluster fitting, we adopt an M31 distance modulus of 24.47, a binary fraction of 0.35, a \citet{Kroupa01} IMF for masses from 0.15 to 120 \solmass, a Milky Way dust attenuation curve ($R_{V}$=3.1), and stellar models from the Padova group \citep{Marigo08} that include updated low-mass asymptotic giant branch tracks \citep{Girardi10}.  We limit the metallicity range to $-0.2 < [M/H] < 0.1$, matching $\sim$$Z_{\sun}$ present day gas phase metallicity observations within M31 \citep{Zurita12, Sanders12}.  A small variation in metallicity is allowed to provide systematic flexibility in the shape and location of the isochrones; metallicity is treated as a nuisance parameter and marginalized over when calculating constraints on the parameters of interest: age, mass, and dust attenuation.

We fit CMDs composed of stars that lie within a cluster's photometric aperture ($R_{\rm ap}$) using radii tabulated in \citet{Johnson15_AP}.
These aperture radii are typically three times the cluster half-light radius. We assume that all cluster members are contained within this radius and make no correction for mass that lies outside the photometric aperture.  We characterize the underlying non-cluster background population using stars that lie in an annulus between $\sim$1.2-3.2 $R_{\rm ap}$, which spans an area 10$\times$ the size of the cluster aperture.  We perform 5$\times$10$^4$ ASTs for each cluster to ensure accurate characterization of photometric completeness and scatter as a function of CMD position and cluster radius.  Input positions for cluster ASTs are distributed radially according to the cluster's luminosity profile, ensuring we derive cluster-integrated photometric properties that accurately reflect the range of conditions in the CMD extraction region.

We compute CMD fits for a grid of age and dust attenuation ($A_V$) values, and obtain mass determinations from the best-fit CMD model scaling at each grid point.  We use relative likelihoods derived across the age-attenuation grid to obtain marginalized probability distribution functions (PDFs) for each of these parameters.  We adopt the age, $A_{V}$, and mass of the best fit model and assign uncertainties to these values based on 16th and 84th percentiles of the marginalized 1D PDFs.  We note that the masses quoted here are initial cluster masses, which are unaffected by gradual mass loss due to stellar evolution.  These cluster masses are appropriate for computing $\Gamma$ because they match the initial masses of the total stellar populations that we derive from SFHs.

\begin{figure}
\centering
\includegraphics[scale=0.6]{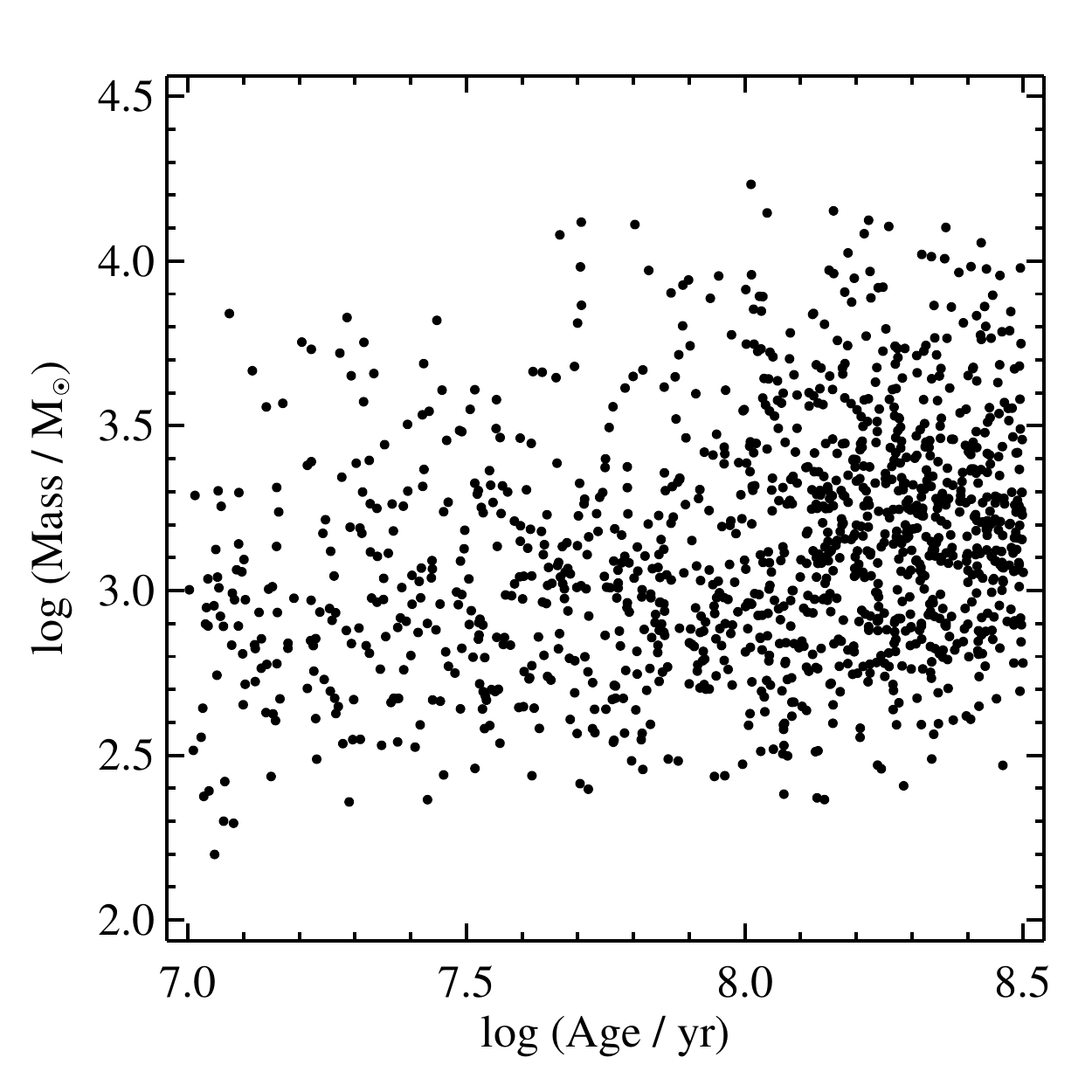}
\caption[Age-mass Distribution for Young PHAT Cluster Sample]{The age-mass distribution for 1249 PHAT/AP clusters in the 10--300 Myr age range.  Random deviations of 0--0.1 dex in age are added to the 0.1 dex grid results to aid visibility.  The increasing density of data points towards larger logarithmic age suggests a uniform linear distribution of cluster ages, as expected under assumptions of a constant formation rate and negligible cluster destruction.}
\label{fig_agemass}
\end{figure}

The fitting results identify 1249 clusters with ages between 10--300 Myr that range in mass from 300 to 20,000 \solmass; the age-mass distribution of the sample is shown in Figure \ref{fig_agemass}.  A notable feature of this plot is the increasing density of data points towards larger logarithmic age, as expected for clusters that are distributed uniformly in linear age.  The sample's age distribution is consistent with a near-constant formation history with little or no cluster destruction, in agreement with initial PHAT results presented in \citet{Fouesneau14}.  The median age uncertainty is 0.2 dex and the median mass uncertainty is 0.04 dex.  We adopt a 0.04 dex (10\%) minimum mass uncertainty for all clusters, reflecting limits in precision due to systematic uncertainties.

We report the ages and masses of the 1249 young clusters analyzed in Appendix \ref{appendix_cat}.  We note that these results are a subsample of the full set of determinations which will be presented in a subsequent paper (A. Seth et~al., in preparation).  This paper will demonstrate the reliability of our cluster CMD fitting using synthetic cluster tests, and compare the CMD-based results to those derived from integrated light fitting.

With ages and masses in hand, we can check for contamination from unbound associations within the 10--300 Myr cluster sample.  Following \citet{Gieles11}, we calculate the ratio of cluster age to crossing time, $\Pi$, using age and mass determinations derived above and photometric half-light radii (equivalent to effective radius, $R_{\rm eff}$) from the AP catalog \citep{Johnson15_AP} to compute $T_{\rm cross}$:
\begin{equation}
T_{\mathrm{cross}} = 10 \left( \frac{R_{\mathrm{eff}}^3}{Gm} \right) ^{1/2}.
\end{equation}
As discussed in Section \ref{definitions}, long-lived gravitationally bound clusters should retain short crossing times as their ages increase, and thus should have $\Pi > 1$ at ages $\ge$10 Myr.

We find that only 33 of the 10--300 Myr clusters have values of $\Pi < 2$ out of 1249 total sample members, a majority of which lie at the 10 Myr young age limit.  This result suggests that contamination from associations is small, even when adopting a liberal threshold for classification (compared to the canonical $\Pi$=1 limit); these candidate associations together make up only 4\% of the total cluster mass in the 10--100 Myr age bin.  Due to the ambiguity in defining a distinct threshold between clusters and associations based on observed $\Pi$ values, and the small effect that excluding these objects would have on the final result, we opt to retain the full cluster sample and make no selection based on $\Pi$.  The small fraction of possible contaminates suggests that adopting a minimum age of 10 Myr for our $\Gamma$ analysis already successfully removed any significant population of potentially unbound stellar associations.

We calculate total observed cluster masses, $M_{\rm cl, obs}$, by summing best fit cluster masses from each of the seven analysis regions that fall within the 10--100 Myr and 100--300 Myr age bins.  We derive uncertainties on these quantities by adding individual cluster mass uncertainties in quadrature.  Region-by-region results are provided in Table \ref{tbl_obs}.

\subsection{Star Formation Histories} \label{analysis_sfh}

The second ingredient for calculating $\Gamma$ is a measurement of the total stellar mass formed during the same age interval as the stellar clusters.  We use recent SFHs calculated in \citet{Lewis15} for this purpose.  Here we provide a high-level overview of the analysis and results, and refer the reader to the original paper for complete details.

SFHs were derived from CMDs using the same MATCH software that was used for cluster fitting.  The SFR is allowed to vary as a function of time for full SFH fitting (fit here with 0.1 dex resolution in logarithmic age), in contrast to cluster fitting that adopts the strong assumption of a simple stellar population.  There are two other differences between the technique for computing extended SFHs rather than cluster SSPs.  First, metallicity is allowed to vary, but is restricted to increase with time.  Second, dust attenuation is implemented using a two-parameter top hat model, defined by a minimum attenuation level and a differential spread that is more appropriate for a spatially distributed, multiage field population.  Other than these differences, assumptions for distance modulus, IMF, binary fraction, and stellar evolution models match those used for cluster analysis.

\begin{figure*}
\centering
\includegraphics[scale=0.4]{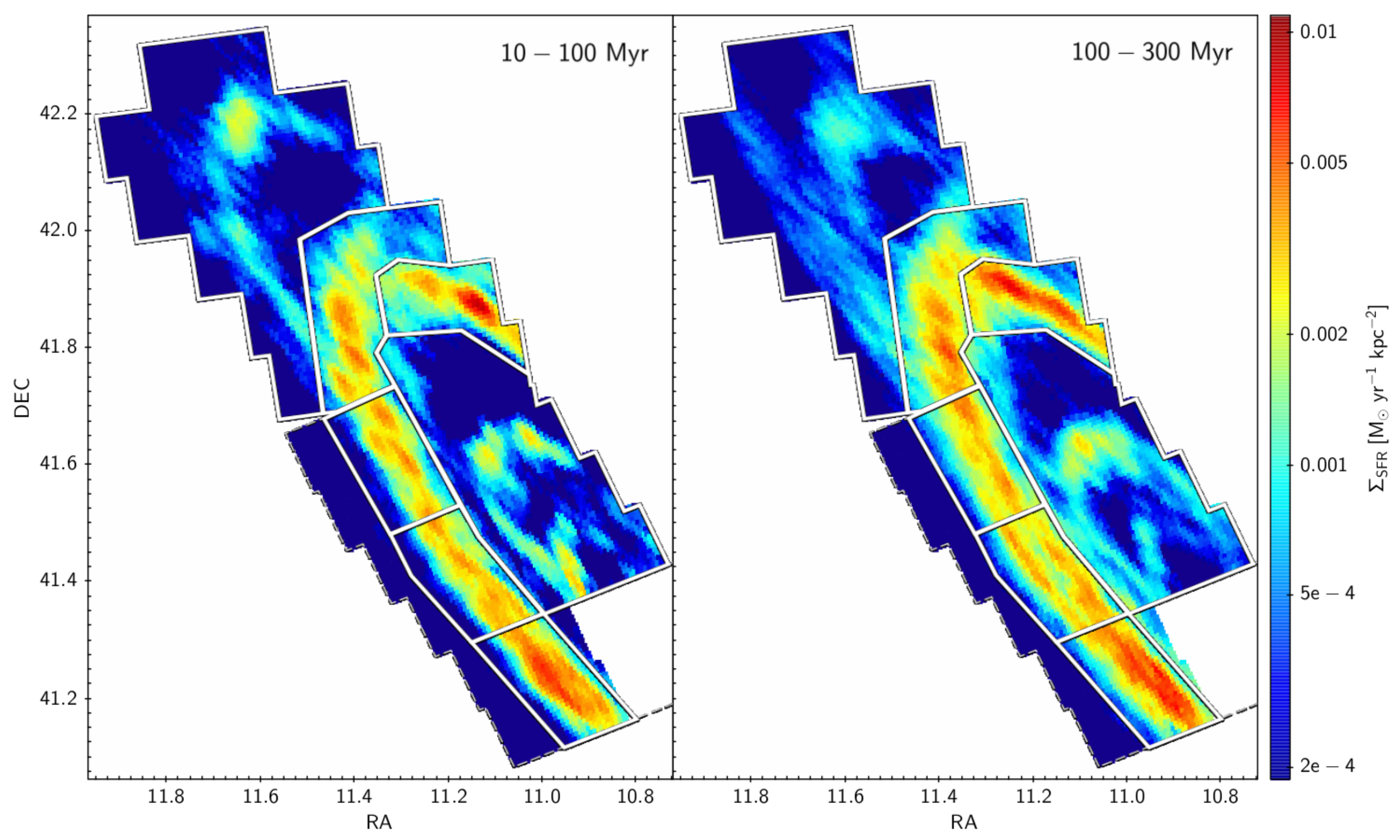}
\caption[\SigSFR\ Maps]{Maps showing \SigSFR\ for 10--100 Myr (left) and 100--300 Myr (right) age bins, which are smoothed with a deprojected 0.5 kpc$^2$ kernel.}
\label{fig_sigsfr}
\end{figure*}

\citet{Lewis15} present SFHs derived on $\sim$100 pc spatial scales for $\sim$9000 individual regions that span the PHAT survey footprint, each measuring 24$\times$27 arcsec.  The SFH for each region were fit using CMDs extracted from the PHAT \texttt{gst} photometry catalogs (described in Section \ref{data_phat}), and $\sim$5$\times$10$^4$ ASTs drawn from a 120$\times$135 arcsec area surrounding the region.  The use of local ASTs ensured that the photometric completeness and scatter adopted by MATCH was appropriate for each region.

Random uncertainties associated with the \citet{Lewis15} SFHs were calculated using a hybrid Monte Carlo (HMC) process \citep{Dolphin13}, producing 10$^4$ posterior samples of SFH parameter values.  The 1$\sigma$ uncertainties are calculated by identifying the region of parameter space with the highest probability density, containing 68\% of the samples.  In addition to these random uncertainties, there are possible sources of systematic uncertainties due to the adopted dust model parameters and the choice of stellar evolution models.  For the purpose of our $\Gamma$ analysis, we ignore both of these sources of uncertainty.  First, the systematic uncertainty due to dust is negligible compared to the random uncertainties.  Second, although there are non-trivial uncertainties and biases associated with adopting a specific set of stellar evolution models \citep[see][]{Dolphin12}, our conclusions are based on relative SFRs and cluster masses that we derive self-consistently using a single set of model assumptions.  Because any systematic offset is shared between the cluster and field results, we also omit this component of uncertainty from the error budget.

We combine best fit SFHs from \citet{Lewis15} spatially within each $\Gamma$ analysis region, and temporally using 10--100 Myr and 100--300 Myr age intervals, to obtain total stellar population masses, $M_{\rm tot}$.  We estimate uncertainties on the integrated mass determinations using a second Monte Carlo sampling analysis.  For each constituent SFH solution, we compute 1000 realizations at full time resolution based on confidence intervals derived from the HMC analysis.  We combine sets of SFH realizations temporally and spatially following the same procedure applied to the best fit results, and define uncertainties based on the scatter in integrated mass among the random samples.  This method will tend to overestimate uncertainties on age-integrated masses due to significant covariance between neighboring age bins at high time resolution.  However, we find that these mass uncertainties are already sufficiently small such that they do not dominate the ultimate $\Gamma$ error budget, and any additional decrease in the total stellar mass uncertainty would have little or no effect on subsequent constraints.  The resulting masses and uncertainties are presented in Table \ref{tbl_obs}.

\subsubsection{Calculating \SigSFR} \label{sigsfr}

We also use the \citet{Lewis15} SFHs to construct \SigSFR\ maps that complement the $M_{\rm tot}$ determinations.  We calculate \SigSFR\ maps using SFRs derived from spatially resolved SFHs integrated over 10--100 Myr and 100--300 Myr age bins, and smooth the results with the same deprojected 0.5 kpc$^2$ measurement kernel applied to the gas surface density maps in Section \ref{data_radio}.

We present \SigSFR\ maps of the PHAT survey region in Figure \ref{fig_sigsfr}.  The spatial distribution of star formation in M31 is highly non-uniform, featuring a prominent 10 kpc star forming ring as well as ring/arm structures in the inner and outer disk.  The OB54 and OB30/31 regions stand out as prominent star forming complexes in the 10 kpc ring, lying at opposite ends of the ring segment observed by the PHAT survey.

We compute characteristic \SigSFR\ values for each of the seven spatial analysis regions by calculating a SFR-weighted average over the set of individual \SigSFR\ values measured in each region.  Analogous to the mass-weighted approach used to calculate characteristic gas surface densities in Section \ref{data_radio}, a SFR-based weighting technique yields characteristic \SigSFR\ values that are minimally affected by the non-uniform spatial distribution of M31's star formation and the specific boundaries used to define the analysis regions. Weighted \SigSFR\ calculations are particularly important for the inner and outer disk regions where star formation takes place within discrete arm/ring structures that have small filling factors.

All previous $\Gamma$ analyses use surface density area normalizations defined simply by the size of the aperture that was used, yielding area-weighted \SigSFR\ estimates.  In the case of non-uniform spatial distributions and small filling factors for star formation activity, the resulting surface densities are sensitive to aperture size.  The inclusion of large areas with relatively low SFRs drives area-weighted \SigSFR\ estimates to artificially low values, even in the case where all the star formation within a given region takes place in a small, high \SigSFR\ subregion.  While the adoption of area-normalized surface densities by previous studies was often out of necessity (e.g., when SFR estimates were not available at higher spatial resolution), these area-weighted values are susceptible to biases, particularly in the case of non-uniform, clumpy spatial distributions.  In contrast, SFR-weighted averaging better characterizes the kpc-scale surface densities at which most of the star formation takes place.  We report these weighted mean \SigSFR\ values for each analysis region in Table \ref{tbl_obs}, and perform a detailed comparison of SFR-averaged and area-averaged surface densities in Appendix \ref{appendix_sigma}.


\subsection{Deriving $\Gamma$} \label{analysis_gamma}

We combine cluster masses with total stellar masses derived from SFH analysis to determine the fraction of stellar mass born in long-lived star clusters, $\Gamma$, over 10--100 Myr and 100--300 Myr age ranges.  Here we introduce a forward modeling approach for transforming measurements of cluster mass and total stellar mass into $\Gamma$ constraints, accounting for unobserved cluster mass and discrete sampling of the cluster mass function.

Our methodology uses two primary observational inputs: the total observed cluster mass ($M_{\rm cl,obs}$) and the total coeval stellar mass ($M_{\rm tot}$).  However, note that $\Gamma$ is defined in terms of total cluster mass ($M_{\rm cl,tot}$), not just the observed cluster mass total ($M_{\rm cl,obs}$).  As part of the modeling, we transform between $M_{\rm cl,obs}$ and $M_{\rm cl,tot}$ using the completeness functions described in Section \ref{clustercat} and assuming a cluster mass function shape.  In this work, we adopt a Schechter function form for the cluster mass function,
\begin{equation}
\mathrm{d}N/\mathrm{d}m \propto m^{\alpha} \exp(m/m_c)
\end{equation}
over the range 10$^2<m/M_{\sun}<10^7$, where $\alpha$ is the low mass slope and $m_c$ is the characteristic cluster mass that sets the position of the exponential turnover.  We adopt a minimum cluster mass of 100 \solmass\ due to the short evolutionary timescales for less massive clusters that would lead to their rapid destruction \citep[$<$10 Myr;][]{Moeckel12} and to provide consistency with previous $\Gamma$ studies.  We adopt $\alpha$=$-2$ and $m_c$=8.5$^{+2.8}_{-1.8} \times 10^3$ \solmass, based on mass function fitting of the PHAT young cluster sample (L. C. Johnson et~al. 2016, in preparation).

We note that the adopted values of $\alpha$, $m_c$, and the minimum cluster mass affect the scaling of our $\Gamma$ measurements.  As an example, our use of a Schechter mass function and a relatively small $m_c$ value yields $\Gamma$ values that are systematically larger by a factor of 1.2--1.5 than if we had adopted a truncated single power law model with a maximum mass between 0.6--4$\times$10$^5$ \solmass, as assumed by \citet{Adamo15} in their analysis of M83.  While it is useful to understand how mass function assumptions factor into the $\Gamma$ results, we are confident in the appropriateness of the $\alpha$ and $m_c$ values adopted here, which are based on Schechter function modeling derived explicitly for the PHAT cluster sample analyzed here.

Our modeling also accounts for the discrete sampling of the cluster mass function and its effect on $\Gamma$ constraints.  Briefly, discrete sampling of the cluster mass function acts as a source of statistical noise when modeling $M_{\rm cl,obs}$ values.  Even when intrinsic values of $\Gamma$ and $M_{\rm tot}$ are held constant, stochastic variations in the distribution of individual cluster masses can cause predictions of $M_{\rm cl,obs}$ to vary.  This effect dominates the error budget in this study's $\Gamma$ determinations due to our tight constraints on $M_{\rm cl,obs}$ and $M_{\rm tot}$ and the limited number of clusters contained in each region per age bin ($\sim$80--100).

To account for stochastic variations in the cluster mass function in our calculations, we formulate a model that predicts $M_{\rm cl,obs}$ as a function of $\Gamma$, $M_{\rm tot}$, and several other input parameters.  The model begins by calculating a total cluster mass, $M_{\rm cl, tot}$, from the input parameters $\Gamma$ and $M_{\rm tot}$.  Next, a random seed value ($X$) is used to initiate a random draw of discrete cluster masses from the Schechter mass function described above, yielding a simulated cluster sample. Finally, using detection probabilities assigned to each of the clusters according to empirically-derived catalog completeness functions (defined in terms of $m_{\rm lim}$, $s_{\rm lim}$; see Section \ref{clustercat}), we simulate an observed subset of clusters and sum the masses of the ``detected'' objects to obtain a prediction for the observed cluster mass, $\widehat{M}_{\rm cl, obs}$.

The resulting probability distributions for $M_{\rm cl, obs}$ are well described by a Gaussian function, therefore we use the following likelihood function to quantify the agreement between observed and predicted quantities:
\begin{equation}
P(M_{\rm cl, obs} | \theta ) = \frac{1}{\sqrt{2 \pi}\sigma_{\rm cl, obs}} \exp \left[ - \frac{(M_{\rm cl, obs}-\widehat{M}_{\rm cl, obs}(\theta))^2}{2\sigma_{\rm cl, obs}^2} \right] ,
\end{equation}
where $\theta$ represents the set of model parameters, $\{\Gamma, M_{\rm tot}, \alpha, m_c, m_{\rm lim}, s_{\rm lim}, X \}$, and $\sigma_{\rm cl, obs}$ represents the uncertainty in the observed cluster mass.  Using Bayes's theorem, we express the posterior probability of the model parameters in terms of the likelihood function:
\begin{equation}
P(\theta | M_{\rm cl,obs}) \propto P(M_{\rm cl,obs} | \theta) P(\theta).
\end{equation}
The $P(\theta)$ term represents the priors on the model parameters.  We adopt a flat prior for $\Gamma$ ($0 \le \Gamma \le 1$) and use region-specific Gaussians derived from the SFHs (mean and $\sigma$ values are listed in Table \ref{tbl_obs}) to define the prior on $M_{\rm tot}$.  We use fixed region-specific values for the completeness function parameters $m_{\rm lim}$ and $s_{\rm lim}$, as listed in Table \ref{tbl_obs}.  Finally, we use a fixed value of $\alpha$=$-2$ and a Gaussian with mean of 3.93 and $\sigma$ of 0.12 as a prior on $\log m_c$ across all regions.

We use a Markov Chain Monte Carlo (MCMC) technique to efficiently sample the posterior probability distribution.  Specifically, we use the \texttt{emcee} package \citep{ForemanMackey13} and its implementation of an affine invariant ensemble sampler from \citet{GoodmanWeare10}.  For our fitting, we use 400 walkers, each producing 2000 step chains, of which we discard the first 100 burn-in steps.  After completing the MCMC computation, we compute a marginalized posterior probability distribution for $\Gamma$, $P(\Gamma | M_{\rm cl,obs})$.  We adopt the median value of the distribution as our primary $\Gamma$ result and report the 16th to 84th percentile range as our 1$\sigma$ confidence interval.

Throughout this paper we assume that cluster dissolution has a negligible effect over the adopted age range.  As a result, we make no adjustment to the total cluster mass other than the mass function extrapolation down to a minimum cluster mass of 100 \solmass\ to transform from $M_{\rm cl,obs}$ to $M_{\rm cl,tot}$.  If cluster disruption were significant, the true value of $\Gamma$ would be larger than the result we obtain.  We discuss the justification for this assumption in detail in Section \ref{discuss_dissolution}.

We conclude here with a brief review of the advantages of our probabilistic modeling approach for calculating $\Gamma$ constraints.  Our main motivation for pursuing probabilistic fitting is its natural ability to derive robust confidence intervals for $\Gamma$, the lack of which has been a shortcoming of previous work.  We note, however, that recent studies have improved in this regard.  For example, statistical variations due to discrete cluster mass function sampling were accounted for by \citet{Ryon14} and \citet{Adamo15}, as well as by \citet{Cook12} in a limited sense.  Within a probabilistic framework, we self-consistently combine constraints on individual input parameters while simultaneously accounting for extrapolation and stochastic sampling of the cluster mass function.  Finally, our forward modeling approach allows a straightforward way to incorporate empirically-derived cluster completeness limits, allowing us to use the entire observed population instead of limiting cluster analysis via conservative lower mass cutoffs like previous $\Gamma$ studies \citep[e.g.,][]{Adamo15}.

\section{Results} \label{results}

In this section, we calculate $\Gamma$ for the PHAT clusters using techniques and results from the previous section and compare to theoretical predictions.  We present observational results in Section \ref{results_gamma} and model predictions from \citet{Kruijssen12} in Section \ref{results_theory}.

\subsection{$\Gamma$ Results} \label{results_gamma}

\begin{figure*}
\centering
\includegraphics[scale=0.72]{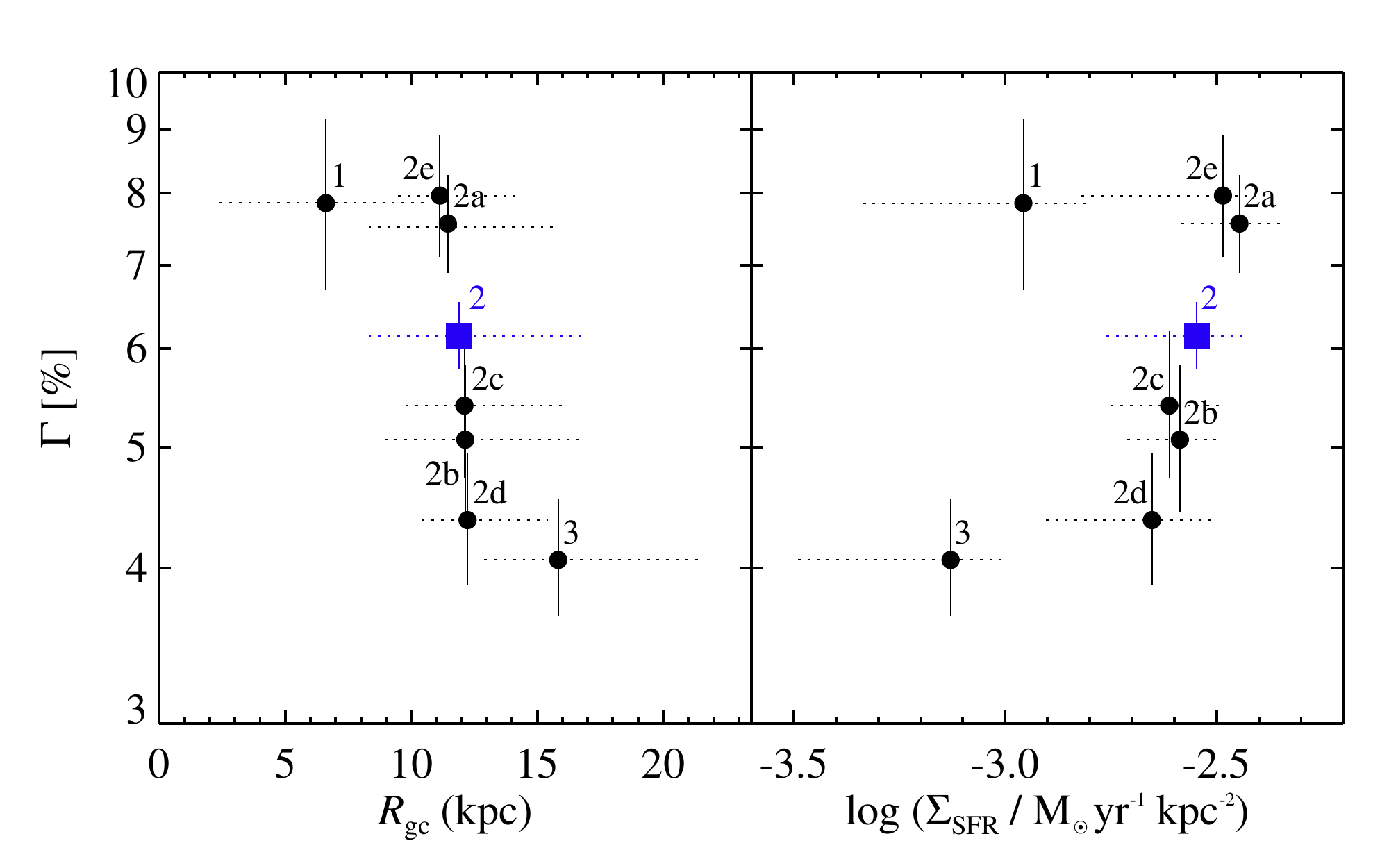}
\caption[$\Gamma$ Results: Dependence on $R_{\rm gc}$ and \SigSFR]{$\Gamma$ results for the 10--100 Myr age bin computed for each analysis region (black circles).  We also show an aggregated data point (Region 2; blue square) representing the combined result for the five regions in the 10 kpc star forming ring (Regions 2a--2e; where log \SigSFR\ $> -2.8$, and $10 < R_{\rm gc} < 13$).  Left: Points are plotted at the median value of the $\Gamma$ PDF and the region's SFR-weighted mean $R_{\rm gc}$.  The solid vertical bars represent the 16th--84th percentile range of the $\Gamma$ PDF, and the dotted horizontal bars represent the full $R_{\rm gc}$ range of each analysis region.  Right: $\Gamma$ results are plotted as in left panel, but now as a function of \SigSFR.  Dotted horizontal bars represent the 25th--75th percentile range of the region's \SigSFR\ distribution.  Uncertainties on the mean \SigSFR\ values are on the order of the markers.}
\label{fig_gam}
\end{figure*}

We derive the probability distribution function of $\Gamma$ in the 10--100 Myr age range for each of the spatial analysis regions, plot the results in Figure \ref{fig_gam}, and report our findings in Table \ref{tbl_gamma}.  We show that $\Gamma$ varies between 4--8\% across the PHAT survey region in M31.  Only a small fraction ($<$10\%) of the stellar mass formed in the last 100 Myr was bound into star clusters.  Low cluster formation efficiencies were expected given Andromeda's relatively quiescent star formation activity and the empirical correlation between $\Gamma$ and star formation intensity (or \SigSFR) established by previous observations.  The $\Gamma$ uncertainties benefit from high precision cluster and total stellar mass determinations, low mass completeness limits for cluster catalogs that reduce extrapolations, and wide accessible age ranges made available by CMD-based SFHs.  In the end, uncertainties on the cluster formation efficiencies are dominated by the contribution from stochastic sampling of the cluster mass function.

We observe statistically significant variations in cluster formation efficiency among the analysis regions, which shows the richness of behavior captured by spatially resolved studies of $\Gamma$ that is otherwise averaged out in galaxy-scale analyses.  In Figure \ref{fig_gam}, we examine how these measured differences in $\Gamma$ correlate with \SigSFR\ and galactocentric radius ($R_{\rm gc}$) in an effort to understand what drives these variations.

In the left panel, we observe that $\Gamma$ varies in a broad sense with galactocentric radius ($R_{\rm gc}$), with cluster formation efficiencies in the outer disk region that are a factor of $\sim$2 lower than in the inner disk, and a mean efficiency for the 10 kpc ring that sits at an intermediate value.  Yet, Figure \ref{fig_gam} also shows that the behavior of $\Gamma$ in M31 is more complex than a simple radial trend.  Within the 10 kpc ring we find variations in $\Gamma$ that span the full 4--8\% range in spite of all five regions lying at approximately the same $R_{\rm gc}$.

The right panel of Figure \ref{fig_gam} seems to show even less of a systematic trend between $\Gamma$ and \SigSFR.  Again, regions from the 10 kpc ring form a tight, steep sequence of points in the plot, but inner and outer disk data points lie parallel to this sequence at lower values of \SigSFR.  It is particularly notable that the inner disk region shows such high $\Gamma$, equaling values found in the two most intense star forming regions in the 10 kpc ring, yet it shows such a low \SigSFR.  

\begin{figure}
\centering
\includegraphics[scale=0.65]{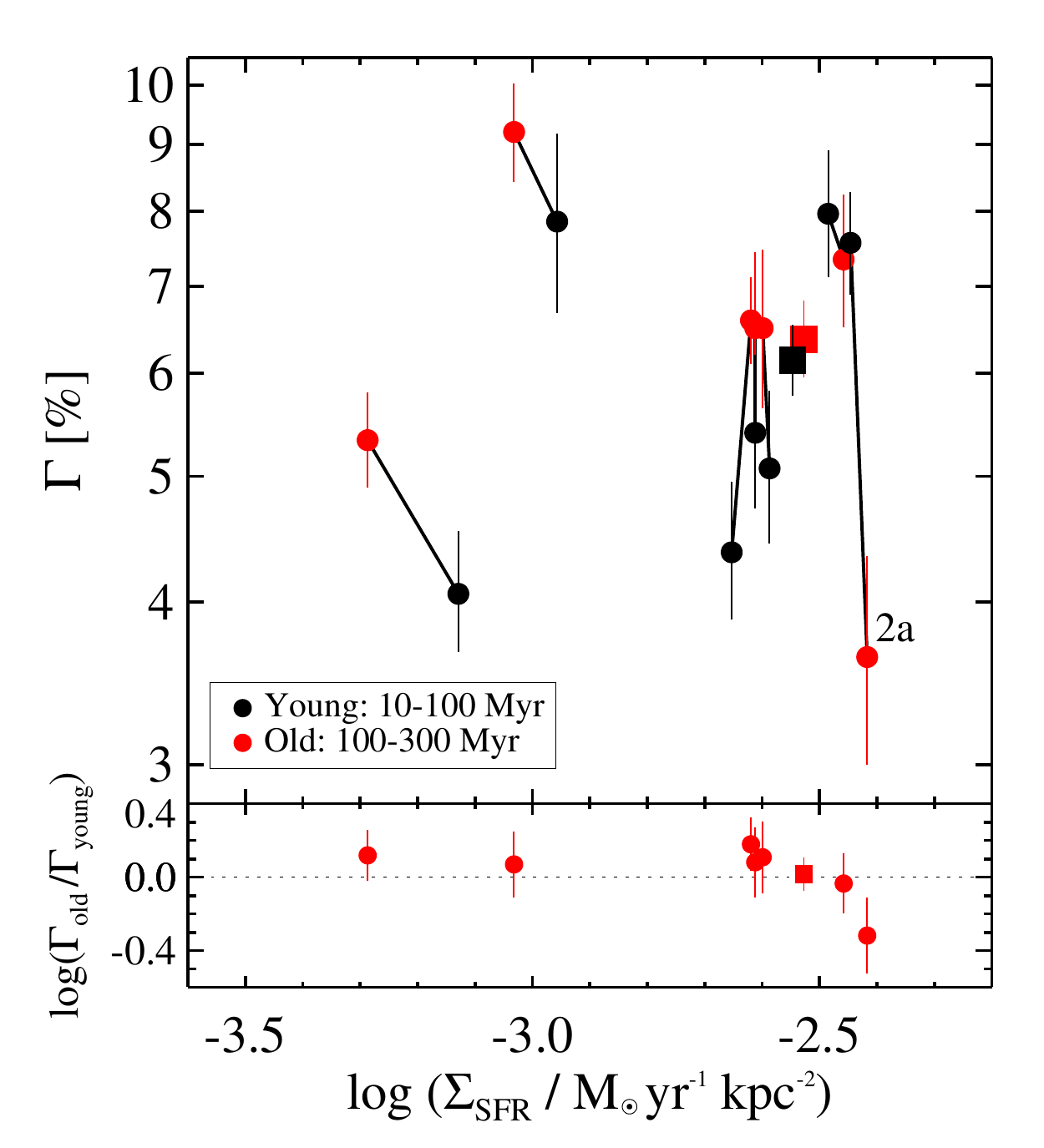}
\caption[$\Gamma$ Results: Comparison with Age]{$\Gamma$ results as a function of age.  Comparing derived quantities for the analysis regions measured over two age ranges, 10--100 Myr (young; black points) and 100--300 Myr (old; red points).  The square points correspond to the combined 10 kpc ring results.  Top: Black lines connect $\Gamma$ data points for the same region.  Bottom: Logarithmic $\Gamma$ residuals between the 100--300 Myr and 10--100 Myr age bins.  The data shows an average, factor of 1.3 ($\sim$0.1 dex) difference between the two age bins.  The notable outlier is Region 2a, due to its anomalously low $\Gamma$ measurement in the 100--300 Myr age bin.}
\label{fig_gam_agecomp}
\end{figure}

We also derive $\Gamma$ results for the 100--300 Myr age bin, report these values in Table \ref{tbl_gamma}, and compare regional $\Gamma$ values from the two age ranges in Figure \ref{fig_gam_agecomp}.  We find generally good agreement between the $\Gamma$ values derived for each age range, as shown by the small residual differences between the two age bins plotted in the figure's bottom panel.  On average, $\Gamma$ measurements in the older age bin were larger by a factor of 1.3 ($\sim$0.1 dex).  The consistency between the age ranges provides evidence that cluster dissolution is negligible over the full 10--300 Myr age range.  If significant cluster dissolution was occurring on these timescales, we would expect the 100--300 Myr $\Gamma$ values to lie below their 10--100 Myr values.  The lone exception to the consistency with age is the southern-most subregion in the 10 kpc ring that hosts the OB30/31 star forming complex (Region 2a; the highest \SigSFR\ data point).  There is no clear explanation for the anomalous, low 100--300 Myr $\Gamma$ measurement and the large accompanying age-dependent difference for the region.  We note that this region lies at the leading end of a continuous string of star forming regions on the northeastern portion of the 10 kpc star forming ring, and active star formation has proceeded throughout the region over the full 10--300 Myr age range (see the \SigSFR\ map in Figure \ref{fig_sigsfr}).

\subsection{Theoretical $\Gamma$ Predictions} \label{results_theory}

The theoretical framework presented in \citet{Kruijssen12} makes predictions for the fraction of stellar mass formed in long-lived star clusters.  This model is based on the idea that bound star clusters naturally arise from a hierarchically structured ISM, where clusters form from gas in the high-density tail of a lognormal distribution.  The free-fall time is short in these high gas density regions, allowing time-integrated efficiencies calculated over the total duration of star formation (until it is truncated due to feedback processes or gas exhaustion) to reach high values, increasing the likelihood of star cluster formation.  \citet{Kruijssen12} developed a self-consistent framework that combines: 1) a model of a turbulent ISM within a gaseous disk that obeys hydrostatic equilibrium, 2) a model of star formation that dictates a specific efficiency per free-fall time \citep{Elmegreen02, KrumholzMcKee05}, 3) a model for the efficiency of initial cluster formation, and 4) ``cruel cradle'' tidal destruction of stellar structures during the gas embedded phase ($<$3-5 Myr).

We calculate theoretical $\Gamma$ predictions using code\footnote{We use the ``global'' version of the code that accepts observable inputs, available at \url{http://www.mpa-garching.mpg.de/cfe/}.} published by \citet{Kruijssen12}.  We combine M31 observations from a variety of sources and compute model input parameters for each of the spatial analysis regions, as described below.

In terms of observable inputs, cluster formation efficiency predictions primarily depend on \SigGas\ according to the \citet{Kruijssen12} model.  The model also accepts two other secondary input parameters to characterize environmental conditions of star forming regions: Toomre $Q$ and angular velocity ($\Omega$). Beyond these three observables, there are additional parameters that control the star formation prescription, the state of the gas and GMCs, feedback mechanisms from star formation processes, and the timescales for termination of star formation.  We adopt default choices for most of these parameters, including an \citet{Elmegreen02} star formation prescription that dictates a single fixed star formation efficiency per free-fall time, and a SN-driven feedback prescription.  We only depart from the standard assumptions of \citet{Kruijssen12} in the case of the $\phi_{P}$ parameter.

The $\phi_{P}$ parameter is a dimensionless constant that encodes the relative contribution of stars and gas to the mid-plane pressure ($P_{\rm mp}$) of the galaxy disk.  This factor is defined with respect to $P_{\rm mp}$ in \citet{KrumholzMcKee05} as
\begin{equation} \label{eq_pressure}
P_{\rm mp} = \phi_{P} \frac{\pi}{2} G \Sigma_{\rm gas}^2 ,
\end{equation}
where $\phi_{P}$ is defined as
\begin{equation}
\phi_{P} = \phi_{\rm mp} f_{\rm gas}^{-1}.
\end{equation}
Here, the constants $\phi_{\rm mp}$ and $f_{\rm gas}$ are defined as
\begin{equation}
\phi_{\rm mp} = \frac{\Sigma_{\rm gas}}{\Sigma_{\rm tot}} + \frac{\sigma_{\rm gas}}{\sigma_{*}} \frac{\Sigma_{*}}{\Sigma_{\rm tot}} =
f_{\rm gas} + \frac{\sigma_{\rm gas}}{\sigma_{*}} (1-f_{\rm gas})
\end{equation}
\begin{equation}
f_{\rm gas}=(\Sigma_{\rm gas}/\Sigma_{\rm tot}) ,
\end{equation}
where \SigStar\ is the stellar surface density, $\Sigma_{\rm tot}$$\equiv$\SigGas+\SigStar\ is the total mass surface density, and $\sigma_{*}$ is the velocity dispersion of the stars.  \citet{KrumholzMcKee05} argue in their Appendix A that $\phi_{P}$ should have a constant value of $\sim$3 across a wide range of galactic environments, and \citet{Kruijssen12} adopts this as one of his standard model assumptions.  We note that model predictions for $\Gamma$ increase as $\phi_{P}$ increases.  In M31, we find that $\phi_{P}$ deviates from this assumed value, varies between analysis regions, and significantly affects resulting $\Gamma$ predictions.  As a result, we treat $\phi_{P}$ as an additional input parameter that we vary from region to region.

We compute theoretical $\Gamma$ estimates using region-specific values of \SigGas, $\Omega$, $Q$, and $\phi_{P}$.  To supplement \SigGas\ measurements derived in Section \ref{data_radio}, we calculate $\Omega$ for each analysis region using a SFR-weighted mean $R_{\rm gc}$ and circular velocities motivated by \citet{Corbelli10} rotation curve results: we assume a flat rotation curve with a circular velocity of 250 km s$^{-1}$ for all regions except the inner disk, where we adopt a circular velocity of 200 km s$^{-1}$.  Next, we calculate the Toomre \textit{Q} parameter for the gas disk using the expression
\begin{equation}
Q \equiv \frac{\kappa \sigma _{\textrm{gas}}}{\pi G \Sigma_{\textrm{gas}}} \approx \frac{\sqrt{2}\Omega \sigma_{\textrm{gas}}}{\pi G \Sigma_{\textrm{gas}}}
\end{equation}
where $\kappa$ is the epicyclic frequency, $\sigma_{\textrm{gas}}$ is the 1D velocity dispersion of the gas (as measured in Section \ref{data_radio}), and where the second approximate equality assumes that the rotation curve is flat within the disk region of interest.  Finally, we calculate $\phi_{P}$ using deprojected \SigStar\ determinations from \citet{Tamm12}, a stellar velocity dispersion determination of $\sigma_{*}$=36 km s$^{-1}$ from \citep{Collins11}, and previously described \SigGas\ and $\sigma_{\rm gas}$ constraints.

We present $\Omega$, $Q$, and $\phi_{P}$ values for each analysis region in Table \ref{tbl_ismobs}.  Notably, $\phi_{P}$ values in the 10 kpc ring and outer disk have a mean of 1.6, and the inner disk has a value of 5.6.  These values depart significantly from the default value of $\phi_{P}$=3, showing that the stellar component's contribution to the disk mid-plane pressure is relatively large in the inner disk, and relatively small in the 10 kpc ring and outer disk, with respect to typical galactic conditions.

\subsubsection{Model Results} \label{results_out_theory}

\begin{figure}
\centering
\includegraphics[scale=0.65]{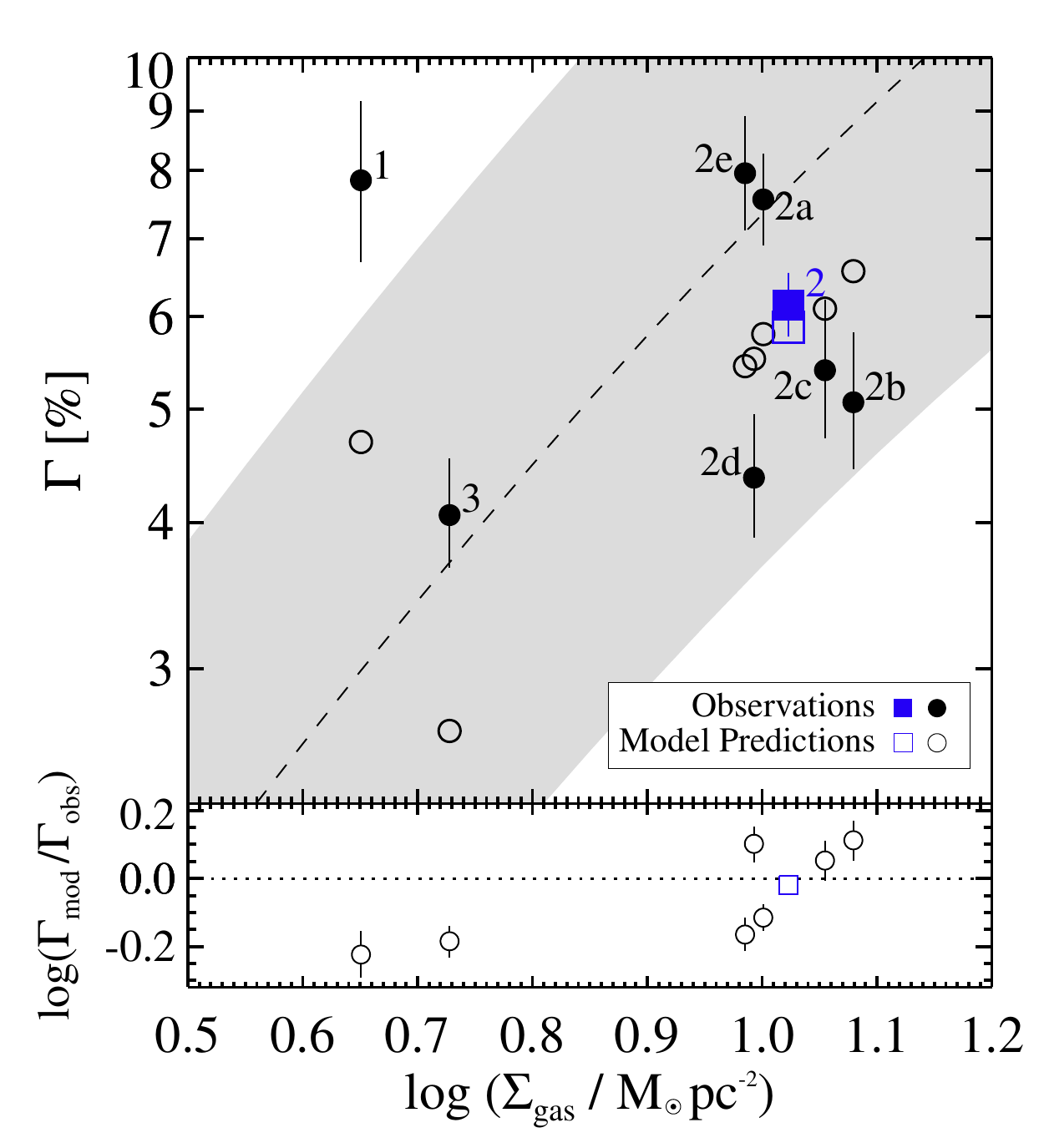}
\caption[Comparison between $\Gamma$ Observations and \citet{Kruijssen12} Model Predictions]{Comparison between $\Gamma$ observations and predictions from the \citet{Kruijssen12} model, presented as a function of present-day \SigGas.  Top: $\Gamma$ observations (filled symbols) and predictions (open symbols) for individual analysis regions are plotted as black circles, and the results for the combined 10 kpc ring are plotted as blue squares.  The dashed curve shows the fiducial \citet{Kruijssen12} $\Gamma$ prediction for typical galactic conditions, and the shaded region represents possible factor of 2 variations around the mean prediction.  The $\Gamma$ observation for the inner disk (Region 1) is high with respect to the fiducial model and its region-specific prediction; we discuss the inner disk region thoroughly in Section \ref{discuss_innerdisk}.  Bottom: Logarithmic $\Gamma$ residuals between region-specific model predictions and observations, showing agreement within a factor of 1.7 ($\sim$0.2 dex) for all analysis regions.}
\label{fig_gam_predict}
\end{figure}

We calculate theoretical $\Gamma$ predictions using observationally derived input parameters from Table \ref{tbl_ismobs}, report these results in Table \ref{tbl_gamma}, and plot the region-by-region predictions along with PHAT measurements in Figure \ref{fig_gam_predict} as a function of \SigGas.  For comparison, we plot the fiducial prediction curve from the \citet{Kruijssen12} model that assumes typical galaxy conditions.  We also plot a shaded region around the curve representing possible factor of 2 variations around this mean relation, accounting for variance in environmental parameters and other model assumptions.  The fiducial prediction increases steadily with gas density over the range of environments found in M31, as expected for low to moderate \SigGas\ environments.  This increasing trend eventually saturates at high \SigGas\ due to ``cruel cradle'' tidal destruction, but rises steadily over the range of environments found in M31.  We also see that the environmental parameters found in M31 can cause individual $\Gamma$ predictions (open circles in Figure \ref{fig_gam_predict}) to differ by a factor of $\sim$1.5 from the fiducial curve.  Specifically, the offsets between fiducial and region-specific predictions seen here (at fixed \SigGas) are primarily due to $\phi_{P}$ values that differ from the default assumption.

At high \SigGas\ values, we observe that the agreement between $\Gamma$ predictions and observations for the integrated 10 kpc ring (Region 2; squares) is very good.  On the other hand, we observe 0.1--0.2 dex scatter between observations and predictions for the five individual 10 kpc regions, as plotted in the bottom panel of Figure \ref{fig_gam_predict}.  This scatter may point to a mismatch between present day \SigGas\ values and the time-averaged properties of the progenitor gas that produced these clusters over the last 100 Myr.  Analyses of the molecular gas in nearby galaxies \citep[e.g.,][]{Kawamura09, Meidt15}, including M31 (L. Beerman et~al., in preparation), have shown that molecular cloud lifetimes are short --- on the order of 20--50 Myr.  Therefore, the cloud population responsible for creating the 10--100 Myr cluster populations are likely no longer in existence due to destructive stellar feedback.  Considering the longevity of the 10 kpc star forming ring \citep[$>$500 Myr;][]{Lewis15}, we can, however, make the assumption that gas properties averaged on ring-integrated spatial scales have remained constant over the past 100 Myr.  Throughout the remainder of the paper, we adopt ring-wide average values for \SigGas\ and other ISM characteristics in the place of region-specific measurements for the five 10 kpc ring analysis regions.

At lower \SigGas\ values, we observe that the $\Gamma$ predictions for the outer disk (Region 3) and inner disk (Region 1) are a factor of 1.5 and 1.7 smaller than the measurements, respectively.  Invoking the same argument used for the 10 kpc ring regions, it is possible that these low predictions are the result of age-dependent scatter in \SigGas.  That said, the $\Gamma$ measurement for the inner disk region is particularly high and may have a physical explanation.  Despite its low \SigGas, the inner disk's relatively high stellar density (\SigStar\ = 94 \solperpc) produces a large $\phi_{P}$, which in turn produces a relatively large $\Gamma$ prediction (4.7\%).  Even with this boost in the predicted value, the $\Gamma$ measurement for the inner disk is still significantly larger than its prediction.  The inner disk observation falls outside the generous factor of 2 range of variation around the fiducial prediction curve, and rivals measurements from the most intense star forming regions in the 10 kpc ring.  We discuss the case of the inner disk region and explore possible explanations for the high cluster formation efficiency in Section \ref{discuss_innerdisk}.

We note that the $\Gamma$ predictions we present here depend on the assumed values of M31 disk properties.  Of these inputs, $\sigma_{*}$ is likely the most uncertain.  We adopt a single survey-wide value of 36 km s$^{-1}$, referencing a measurement of the mass-dominant thin disk component from \citep{Collins11}.  This falls on the low end of the likely range of plausible values, considering the age-dependent 30--90 km s$^{-1}$ range in $\sigma_{*}$ reported in \citep{Dorman15}.  Increasing $\sigma_{*}$ from 36 km/s to 90 km/s would decrease $\phi_{P}$ values by 0.1 dex (0.3 dex for the inner disk).  As a result, $\Gamma$ predictions would decrease by 0.05 dex over most of the survey (0.15 dex for the inner disk) and the discrepancy between model predictions and observations would increase.

\section{Discussion} \label{discuss}

\subsection{Galaxy-wide $\Gamma$ Results} \label{discuss_gamgal}

As discussed in the introduction, a growing body of evidence has revealed that cluster formation efficiency varies as a function of star forming environment.  Beginning with \citet{Goddard10}, numerous studies have measured cluster formation efficiencies at galaxy-integrated scales, revealing a positive correlation between $\Gamma$ and \SigSFR.  Work by \citet{Goddard10}, \citet{Adamo11}, \citet{SilvaVilla11}, and \citet{Cook12} each contribute galaxy-integrated measurements for small samples of galaxies.  Additionally, studies by \citet{Annibali11}, \citet{Baumgardt13}, \citet{Ryon14}, \citet{Lim15}, and \citet{Adamo15} contribute results for individual galaxies.  Together, these studies represent $\Gamma$ measurements for a combined sample of 30 galaxies.  Appendix \ref{appendix_gamlit} provides a detailed discussion about the curation of these results, explaining our preference for measurements that are most similar to our own (e.g., matching age ranges and cluster dissolution assumptions when possible), and includes comments and caveats specific to individual studies.

\begin{figure}
\centering
\includegraphics[scale=0.65]{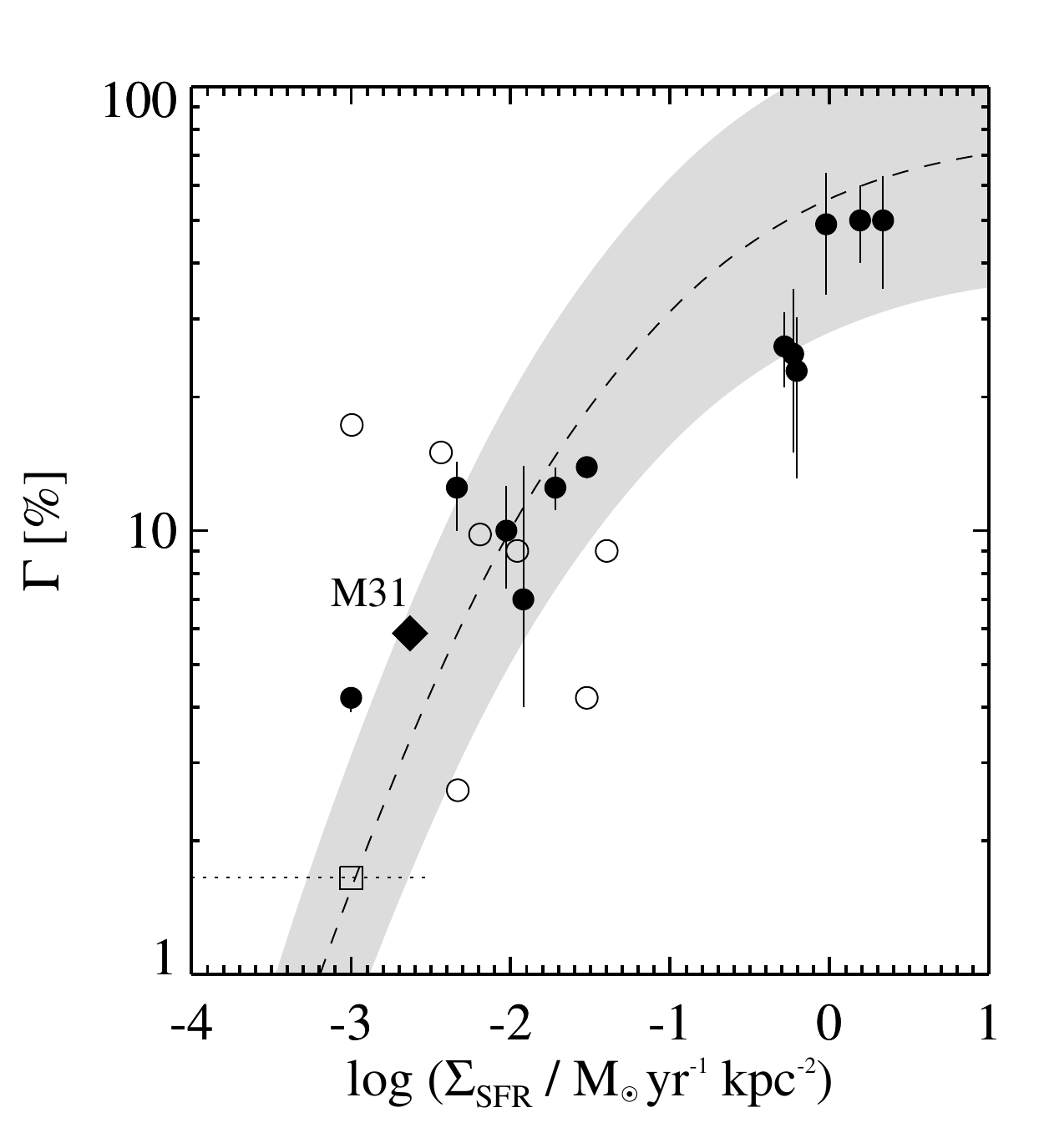}
\caption[Galaxy-integrated $\Gamma$ Results]{The survey-averaged $\Gamma$ measurement for PHAT (diamond) is compared to galaxy-wide results from the literature (see text for references).  Literature results that report uncertainties are plotted using filled symbols, while those without uncertainties are plotted using open symbols.  The binned result from \citet{Cook12} is plotted according to sample-wide average values of $\Gamma$ and \SigSFR\ (open square), and a horizontal dotted line denotes the \SigSFR\ bin width.  The dashed line represents the fiducial $\Gamma$--\SigSFR\ relation for galaxies from \citet{Kruijssen12}, and the shaded region represents a factor of 2 variation around the fiducial relation to account for variations in physical conditions.  This compilation of galaxy-integrated $\Gamma$ measurements shows a positive correlation between $\Gamma$ and \SigSFR, and while there is good overall agreement between observations and the predicted theoretical relation, the observed scatter around the predicted relation is considerable.}
\label{fig_gam_gal}
\end{figure}

We combine data from the seven analysis regions and compute a survey-wide 10--100 Myr $\Gamma$ measurement of 5.9$\pm0.3$\%; we provide a full set of survey-wide results in Tables \ref{tbl_ismobs}, \ref{tbl_obs}, and \ref{tbl_gamma}.  We plot this measurement, the curated set of galaxy-integrated literature values, and compare these results to a predicted $\Gamma$-\SigSFR\ relation from \citet{Kruijssen12} in Figure \ref{fig_gam_gal}.  The fiducial prediction shown here assumes a \SigGas-to-\SigSFR\ conversion that follows from the Schmidt-Kennicutt star formation relation \citep{Kennicutt98} as well as typical environmental parameter values ($Q$=1.5, $\Omega$ derived from empirical relation with \SigGas\ and therefore \SigSFR).  The PHAT survey-wide $\Gamma$ measurement follows the established (but noisy) $\Gamma$-\SigSFR\ trend previously observed, and lies above the predicted relation.

The current compilation of galaxy-wide results shows an empirical trend where $\Gamma$ increases with \SigSFR, but shows significant scatter.  This scatter could be due to physical differences in the observed galaxies, or due to observational heterogeneity and uncertainty.  Differences in analysis techniques, assumptions, and data quality among these heterogeneous $\Gamma$ studies could explain the observed scatter.  For example, authors of these studies differ in their methodology for deriving total SFRs (e.g., resolved stars versus \halpha/FUV luminosity transformations), in the cluster age and mass ranges studied, and in the estimation of uncertainties (including cases where this analysis was not performed).    In particular, we highlight the lack of reported uncertainties, and the underestimation of uncertainties in cases where these values are reported (i.e., not accounting for stochastic sampling of the cluster mass function), as a serious obstacle to differentiating between genuine $\Gamma$ variation and observational scatter.

We can also interpret these results relative to the \citet{Kruijssen12} fiducial curve.  The predicted curve follows the distribution of measurements quite well, and one could try to explain the observed scatter by invoking true physical variations in environmental properties of star formation within and among these galaxies.  \citet{Kruijssen12} is clear in stating that his default $\Gamma$-\SigSFR\ relation represents characteristic predictions of the model, and that one should expect variations of up to a factor of $\sim$2--3 around this relation due to differences in galactic environments (i.e., deviations in \SigGas, Toomre $Q$, $\Omega$, $\phi_{P}$), as well as departures from the standard set of assumptions (e.g., changing the prescription for stellar feedback timescales and mechanisms).  Please see sections 3.4, 7.1, and Appendix C in \citet{Kruijssen12} for further discussion of the sensitivity of $\Gamma$ predictions to variations in input parameters and model assumptions.  We convey uncertainty in the model predictions in Figure \ref{fig_gam_gal} using a shaded region denoting a factor of 2 variation around the fiducial relation.  We therefore conclude that true variations in $\Gamma$ could plausibly produce the scatter in galaxy-scale observations.

We find the overall level of agreement between the observations and model prediction quite impressive, but conclude that the heterogeneous nature of galaxy-integrated $\Gamma$ observations fundamentally limit deeper interpretation of these results. 

\subsection{Spatially Resolved $\Gamma$ Observations and Predictions} \label{discuss_gamres}

The galaxy-integrated observations presented in Figure \ref{fig_gam_gal} provide good evidence for an environmentally dependent cluster formation efficiency, clearly showing that $\Gamma$ increases by an order of magnitude as log (\SigSFR\ / \solperyr\ kpc$^{-2}$) increases from $-3$ to 0. However, averaging over a wide range of star formation environments can hide variations in star cluster formation efficiency occurring on smaller scales within galaxies.  Here we shift our focus to spatially resolved measurements, which allow us to study star cluster formation and its dependence on the physical properties of the ISM in detail.  In this discussion section, we examine observational results from M31 (this work) and M83 \citep{SilvaVilla13, Adamo15}.  We focus on these two galaxies in particular due to the similarity and compatibility of the two analyses \citep[using the 10--50 Myr equal-area region results from][]{Adamo15} and the availability of complementary ISM observations for both galaxies.

To accompany these spatially resolved observations and aid in their interpretation, we derive theoretical $\Gamma$ predictions that are appropriate for this new domain of sub-galactic scale analysis.  While the \citet{Kruijssen12} framework is intrinsically scale independent, the \SigGas-to-\SigSFR\ conversion used to map \SigGas-dependent predictions into an observationally relevant \SigSFR\ parameter space imprints a spatial scale dependence on the existing $\Gamma$--\SigSFR\ relation prediction.  We note that one could avoid the use of a \SigGas-to-\SigSFR\ conversion altogether by directly comparing observations and predictions in the $\Gamma$--\SigGas\ plane, as done in Figure \ref{fig_gam_predict}.  However, due to the variable, age-dependent nature of \SigGas\ (see discussion in Section \ref{results_out_theory}) and the greater availability of \SigSFR\ observations, the $\Gamma$--\SigSFR\ parameter space is an observationally-favorable parameter space for present and future $\Gamma$ analyses.

In Section \ref{sec_newsfl}, we use previously published observations of nearby galaxies to define a new spatially resolved star formation relation (\SigSFR\ $\propto$ \SigGas$^{N}$).  We use the resulting \SigGas-to-\SigSFR\ conversion to derive a new $\Gamma$--\SigSFR\ relation prediction for the \citet{Kruijssen12} theoretical model.  Next, we compare spatially resolved $\Gamma$ measurements from M31 and M83 to the newly derived theoretical predictions in Section \ref{sec_comp}.  Finally, we consider the broader application of our new set of revised $\Gamma$--\SigSFR\ predictions and explore new interpretations of cluster formation efficiency results at high \SigSFR\ in Section \ref{sec_newapp}.

\subsubsection{A Spatially Resolved Star Formation Relation: \SigGas\ versus \SigSFR} \label{sec_newsfl}

When we consider spatially resolved $\Gamma$ constraints, we must adapt assumptions about star formation behavior that were originally calibrated on galaxy-wide scales.  As discussed in the previous section, \citet{Kruijssen12} adopts a Schmidt-Kennicutt star formation relation \citep{Kennicutt98} to convert from \SigGas\ to \SigSFR\ when deriving a fiducial $\Gamma$-\SigSFR\ prediction.  The Schmidt-Kennicutt relation's global $N$=1.4 power law slope and normalization were originally defined using galaxy-integrated measurements of molecular gas dominated systems, sampling moderate to high star formation activity.

In contrast, spatially resolved studies that cover a wide range of star formation environments demonstrate that the relationship between \SigGas\ and \SigSFR\ does not follow a single universal power law.  \citet{Bigiel08} make sub-kpc scale measurements in nearby star forming galaxies and find that \SigSFR\ correlates linearly with \SigGas\ in molecular gas dominated environments, suggesting a constant star formation efficiency in this regime.  However, as the ISM becomes atomic gas dominated at \SigGas\ $\lesssim$10 \solperpc, the star formation relation steepens (and the observed scatter increases), indicating a decline in star formation efficiency as \SigGas\ decreases.  This change in star formation efficiency can be equivalently characterized as a change in the total gas depletion time (\tdep\ $\equiv$ \SigGas\ / \SigSFR), with inefficient star formation at low gas density corresponding to long \tdep.

To account for the observed properties of spatially resolved star formation, we define a new star formation relation based on the \citet{Bigiel08} results.  The new relation captures the change in slope between molecular and atomic gas dominated regimes, and accounts for observed scatter by allowing a range of \SigSFR\ values as a function of \SigGas.  We implement flexibility in the relation by allowing \tdep\ variations that are consistent with these nearby galaxy observations.

\begin{figure}
\centering
\includegraphics[scale=0.48]{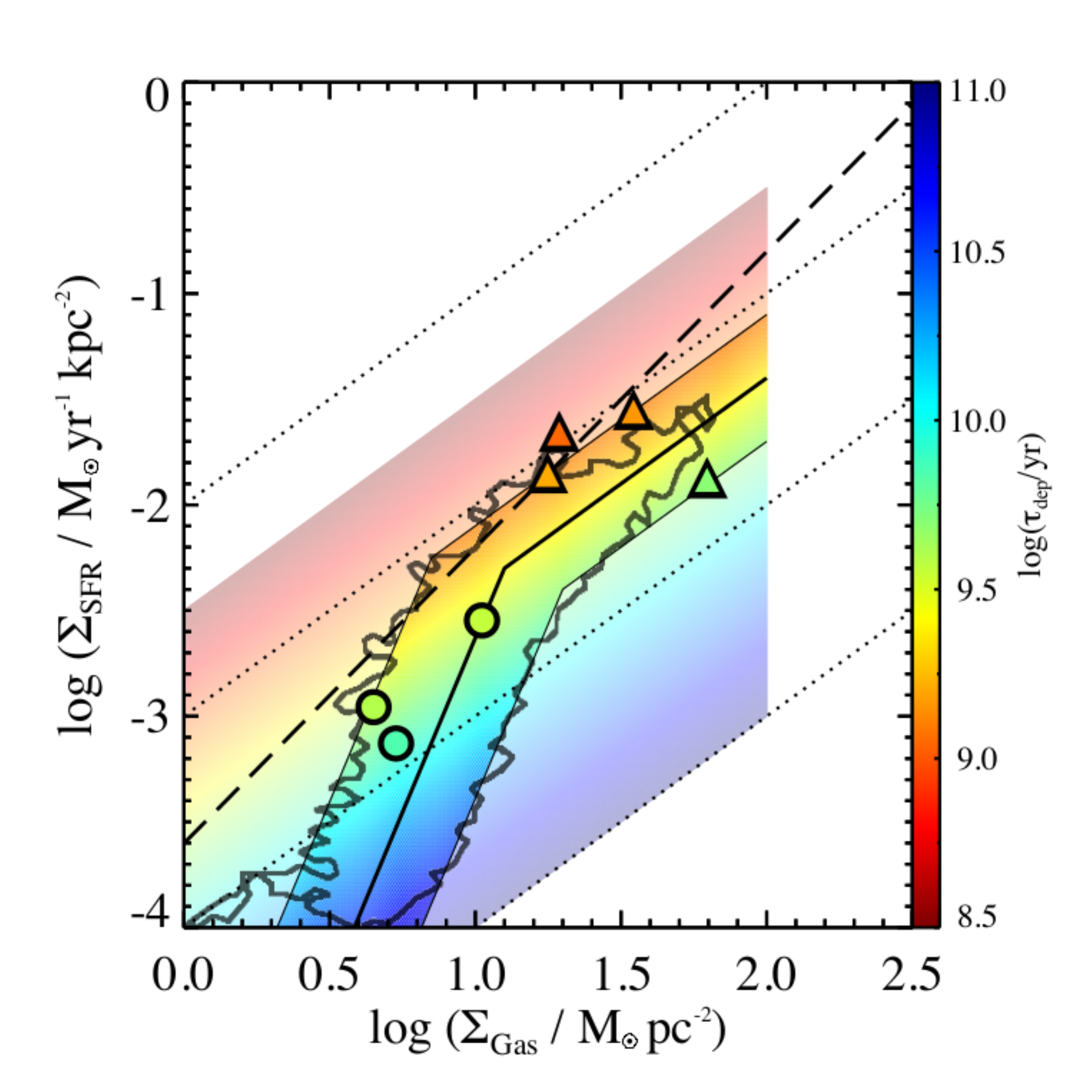}
\caption[\SigGas\ versus \SigSFR: Comparison of Star Formation Relations and Observations]{Comparison between the Schmidt-Kennicutt star formation relation (black dashed line) and spatially resolved observations from \citet{Bigiel08} (gray contour).  We include observations of M31 from this paper (circles; inner disk, outer disk, and 10 kpc ring), and observations from \citet{Adamo15} of M83 (triangles; equal area annuli).  We use a broken power law to characterize the range of \SigSFR\ (and thus \tdep) as a function of \SigGas\ that is consistent with the \citet{Bigiel08} observations; we plot the median two-component star formation relation (thick black line) and its accompanying upper and lower envelopes (thin black lines).  We also plot dotted lines that represent constant \tdep\ for log(yr) of 8, 9, 10, and 11 (from top to bottom), and include a background color gradient encoding \tdep\ values from 8.5 $<$ log(yr) $<$ 11.}
\label{fig_ks}
\end{figure}

We use observations from \citet{Bigiel08} to define the two slope values of the new \SigGas-\SigSFR\ relation, as well as an acceptable range of \tdep\ as a function of \SigGas.  In Figure \ref{fig_ks}, we show that the distribution of \citet{Bigiel08} observations\footnote{We use as reference the distribution of \SigSFR(FUV+24\um) versus \SigGas\ observations, represented by the contiguous portion of the orange contour (denoting a density of 2 samples per 0.05 dex-wide cell) from Figure 8 in \citet{Bigiel08}.  This distribution is shifted by a factor of 1.36 in our work to account for the mass of helium that we include in \SigGas\ that was not included in the original work.}, 
represented by the gray contour, is well-characterized by a two-part star formation relation, where \SigSFR\ $\propto$ \SigGas$^{N}$ with $N$=1 at high \Htwo-dominated gas densities, and $N$=3.3 for low \HI-dominated gas densities.  This behavior differs significantly from the Schmidt-Kennicutt relation, which we plot as a dashed line for comparison.  We characterize the intrinsic scatter using parallel upper and lower thresholds that define an envelope around the median relation, encompassing the \tdep\ variation observed in the data: 0.6 dex for high \SigGas, 1.6 dex at low \SigGas.  Please see Appendix \ref{appendix_sflaw} for a detailed description (i.e., normalizations, limits) of the adopted relations.

We also explore how the M31 and M83 analysis regions compare to the \citet{Bigiel08} observations and our newly derived star formation relation.  We use \SigGas\ and \SigSFR\ measurements listed in Tables \ref{tbl_ismobs} and \ref{tbl_obs} for M31 regions, but remind the reader that we only consider ring-wide \SigGas\ and \tdep\ results for the 10 kpc ring region (see Section \ref{results_out_theory}).  We supplement \SigSFR\ and \SigHtwo\ measurements from \citet{Adamo15} with \SigHI\ measurements from \citet{Bigiel10_M83} to determine \SigGas\  and \tdep\ for the M83 regions.  In Figure \ref{fig_ks}, we observe that three M83 regions and the inner disk data point from M31 (Region 1) lie on the upper envelope of local observations, corresponding to relatively short depletion times and high star formation efficiencies with respect to typical local galaxies.  The innermost annulus in M83 lies near the lower envelope of the \citet{Bigiel08} observations and has a relatively large \tdep, while the 10 kpc ring and outer disk regions from M31 (Regions 2 \& 3) lie on or near the median relation for local star formation observations.

\subsubsection{Comparing Spatially Resolved $\Gamma$ Observations and \tdep-dependent Predictions} \label{sec_comp}

\begin{figure}
\includegraphics[scale=0.48]{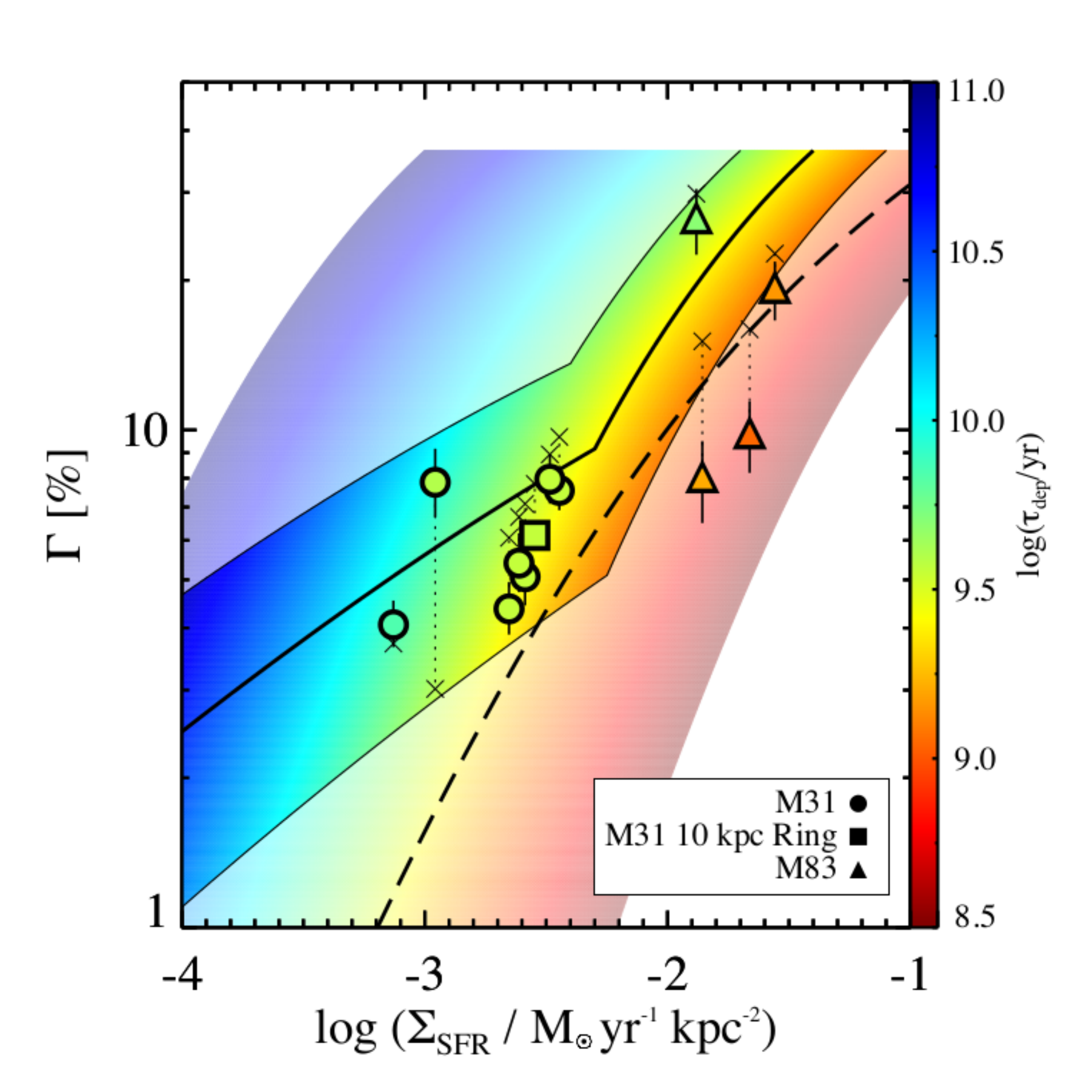}
\caption[Spatially Resolved $\Gamma$ Measurements]{Spatially resolved $\Gamma$ measurements for M31 (circles) and M83 (triangles).  We plot new fiducial $\Gamma$ predictions for the \citet{Kruijssen12} model assuming a spatially resolved star formation relation derived from the \citet{Bigiel08} observations.  Model predictions are plotted for \SigGas\ $<$ 100 \solperpc, color coded by \tdep, which ranges from log(yr) of 8.5 (red) to 11 (violet).  We highlight the portion of model parameter space that is consistent with \tdep\ observations, as defined in Figure \ref{fig_ks}: the thin solid lines represent the upper and lower envelopes to the observed range; the thick solid line represents the median relation.  The original $\Gamma$-\SigSFR\ relation from \citet{Kruijssen12} is plotted as a dashed line.  Data points for M31 and M83 are color coded according to observed \tdep, where good agreement between the model and observations is represented by a color match between the data point and the underlying models.  Positions in the plot representing a region's fiducial $\Gamma$ prediction given observed values of \SigSFR\ and \tdep\ are marked with Xs and connected to the corresponding observations by dotted lines.}
\label{fig_gam_reg}
\end{figure}

We use the newly defined star formation relation from Figure \ref{fig_ks} to compute new $\Gamma$ predictions in terms of \SigSFR\ and \tdep\ in Figure \ref{fig_gam_reg}.  Using the new median star formation relation, we first transform \SigGas-dependent predictions from \citet{Kruijssen12} and derive a new fiducial $\Gamma$-\SigSFR\ relation.  Next, we propagate scatter from the star formation relation by defining upper and lower envelopes around the fiducial $\Gamma$-\SigSFR\ relation, which follow from the upper and lower thresholds defined in Figure \ref{fig_ks}.  The parameter space enclosed by the upper and lower envelopes in Figure \ref{fig_gam_reg} represents the expected range of spatially resolved $\Gamma$ measurements, as predicted by the \citet{Kruijssen12} model. 

We parameterize the variation in $\Gamma$ (as a function of \SigSFR) using \tdep, where the allowed range of \tdep\ is set by the \citet{Bigiel08} observations.  We note that given a specific pair of \SigSFR\ and \tdep\ values (and characteristic values of $Q$, $\Omega$, and $\phi_{P}$), the \citet{Kruijssen12} model uniquely predicts $\Gamma$.  In addition to defining a new $\Gamma$-\SigSFR\ relation, we also produce a generalized set of \tdep-dependent $\Gamma$ predictions as a function of \SigSFR.  We visualize this grid of theoretical predictions using \tdep-based color coding in Figure \ref{fig_gam_reg}.

The new spatially resolved star formation relation imprints a break in the predicted $\Gamma$-\SigSFR\ relation, representing the transition from \Htwo-dominated to \HI-dominated star forming environments.  In Figure \ref{fig_gam_reg}, we show that the predicted relation flattens at low \SigSFR\ due to the dramatic increase in \tdep\ at low \SigGas.  In contrast to fiducial predictions from \citet{Kruijssen12}, we expect low density environments with $-4 <$ log (\SigSFR\ / \solperyr\ kpc$^{-2}$) $< -3$ to form a small percentage of their stars (1--5\%) in long-lived star clusters as opposed to the negligible fraction ($<$1\%) predicted by the steeply declining fiducial $\Gamma$ relation (dashed line).  

In addition to analyzing the behavior of the new fiducial $\Gamma$-\SigSFR\ relation in Figure \ref{fig_gam_reg}, here we highlight and explain two notable trends that emerge from the grid of $\Gamma$ predictions.  First, we find that predicted $\Gamma$-\SigSFR\ relations at fixed values of \tdep\ (curved bands of constant color) are quite steep.  The steepness of these relations reflect the slopes of the underlying star formation relation: $N$=1.0 for the constant \tdep\ case producing a steep $\Gamma$ relation, compared to $N$=1.4 (standard Schmidt-Kennicutt slope) assumed for the shallower fiducial $\Gamma$ relation (dashed line).  Second, we observe that as \tdep\ decreases and star formation efficiency increases, the predicted $\Gamma$ relation moves to the right in Figure \ref{fig_gam_reg} towards higher \SigSFR\ values. We will use these general properties of theoretical $\Gamma$ predictions to help interpret the distribution of current observations in Section \ref{sec_newapp}.

With appropriate model relations in hand, we compare the spatially resolved M31 and M83 $\Gamma$ observations to theoretical predictions in Figure \ref{fig_gam_reg}.  The availability of \tdep\ measurements for these analysis regions (represented by the color coding assigned to each point in Figure \ref{fig_ks}) allows us to individually evaluate the agreement between observations and theoretical predictions in all three relevant parameters ($\Gamma$, \SigSFR, and \tdep).  In the case of good agreement, we expect the data points in Figure \ref{fig_gam_reg} to match the color of the models located at the same position in the plot.  As an additional aid, we plot vertical dotted lines that connect $\Gamma$ observations to the model grid point representing the \tdep\ and \SigSFR\ measurements for each region, representing the offset between observed and predicted cluster formation efficiencies.

We observe that the new \tdep-dependent fiducial $\Gamma$ relation describes the combined M31/M83 dataset very well and represents a significant improvement over the original \citet{Kruijssen12} fiducial relation.  The flattening of the predicted relation at low \SigSFR, due to the increase in \tdep\ at low \SigGas, eliminates previous discrepancies between M31 observations and model predictions.

We also find generally good agreement between the color-coded observations and the underlying $\Gamma$ model grid, reflecting consistency with theoretical predictions in all three parameters: $\Gamma$, \SigSFR, and \tdep.  
In particular, the tight sequence of data points formed by M31's 10 kpc ring regions demonstrates the success of \tdep-dependent $\Gamma$ modeling.  We expect a set of observations with the same \tdep\ to follow a steep line of constant color.  The 10 kpc ring regions, which share a common \tdep, fulfill this prediction accurately by tracing a steep, monochromatic sequence of models in Figure \ref{fig_gam_reg}.

In contrast to cases of excellent consistency, two of the M83 observations lie 2--3$\sigma$ below their $\Gamma$ predictions.  We note, however, that \citet{Adamo15} use a conservative cluster catalog selection criteria (rejecting questionable ``Class 2'' candidates) and state that their reported values could be low due to this cut.  The only seriously discrepant observation is the inner disk region of M31; we will discuss this region in detail in Section \ref{discuss_innerdisk}.

One final point of discussion concerns our use of default model parameter values from \citet{Kruijssen12}.  As we previously discussed in Section \ref{discuss_gamgal} with regard to the fiducial $\Gamma$-\SigSFR\ relation, the new \tdep-dependent $\Gamma$ relation presented here is based on a set of assumptions that describe typical conditions in star forming galaxies.  Unlike the analysis presented in Section \ref{results_theory} that uses region-specific model inputs to calculate specific, detailed $\Gamma$ predictions, general-use relations must adopt canonical sets of input parameter values.  Variations among galactic environmental parameters (i.e., \SigGas, Toomre $Q$, $\Omega$, $\phi_{P}$) or modifications to the standard set of assumptions (e.g., changing the prescription for stellar feedback timescales and mechanisms) could produce variations of up to a factor of 2--3 in $\Gamma$ predictions.  Although poor input parameter assumptions could still produce excess scatter between $\Gamma$ predictions and observations, the new \tdep-dependent modeling provides an effective explanation for a substantial fraction of the dispersion among $\Gamma$ measurements.

In conclusion, we find that the agreement between spatially resolved observations and fiducial \citet{Kruijssen12} model predictions in the $\Gamma$-\SigSFR\ plane greatly improves at low \SigSFR\ when we use a multi-component star formation relation that accounts for variations in \tdep\ as a function of \SigGas.  The resulting \tdep\ variations also serve as a plausible explanation for the scatter in the observed $\Gamma$-\SigSFR\ distribution. 

\subsubsection{A Starburst $\Gamma$ Relation: Short \tdep\ or Radiative Feedback?} \label{sec_newapp}

To close discussion of the \tdep-dependent $\Gamma$ relations, we expand beyond normal galaxies to consider more intense, starburst environments.  The non-linear slope of the Schmidt-Kennicutt relation ($N$=1.4) indicates that for galaxy-integrated scales, \tdep\ decreases as \SigGas\ increases.  This result is rather intuitive, suggesting that gas collapses into stars more efficiently at higher densities \citep[though remaining constant per free-fall time;][]{Krumholz12}.  These high star formation efficiencies are found in (U)LIRGs and other starburst galaxies, but also in the dense central regions of otherwise normal galaxies.

\begin{figure}
\includegraphics[scale=0.7]{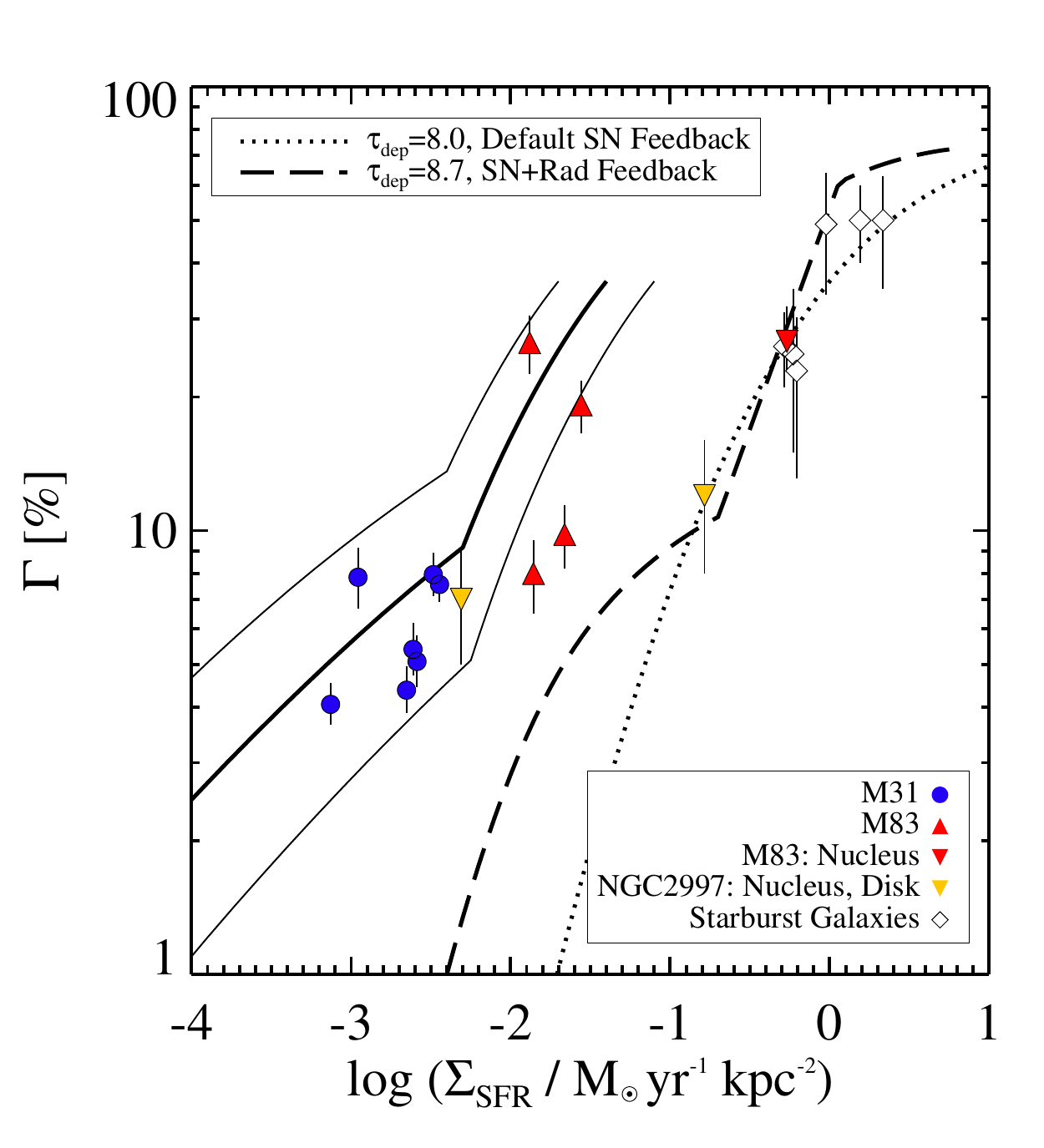}
\caption[Starburst Galaxies and the \tdep-dependent $\Gamma$ Relation]{Reinterpreting starburst $\Gamma$ observations using \tdep-dependent $\Gamma$ relations.  We plot spatially resolved measurements from M31 (circles; this work), M83 \citep[upward triangles;][]{Adamo15,Goddard10}, and NGC2997 \citep[downward triangles;][]{Ryon14}, as well as integrated measurements from six starburst galaxies \citep{Goddard10, Adamo11}.  Theoretical $\Gamma$ relations from \citet{Kruijssen12}: log(\tdep/yr)=8.0 with nominal SN-only feedback (dotted line), and log(\tdep/yr)=8.7 with an alternative SN+Rad combined feedback prescription (dashed line).  The predicted parameter space for spatially resolved $\Gamma$ observations in normal galaxies is represented by solid lines, as in Figure \ref{fig_gam_reg}.  We observe that regions with \SigSFR\ $>$ 0.1 \solperyr\ kpc$^{-2}$ are consistent with our new $\Gamma$ predictions when \tdep\ $\sim$ 100 Myr.  Altering the model's stellar feedback prescription also leads to a satisfactory fit for a longer \tdep\ of $\sim$500 Myr.}
\label{fig_gam_starburst}
\end{figure}

From the current set of observations, three studies have placed constraints on the fraction of stellar mass born in long-lived clusters within starburst environments (\SigSFR\ $>$ 0.1 \solperyr\ kpc$^{-2}$): \citet{Goddard10} analyzed NGC3256 and the nuclear region of M83, \citet{Adamo11} analyzed a sample of five blue compact galaxies, and \citet{Ryon14} analyzed the nuclear region of NGC2997.  We plot these observations along with the spatially resolved M31 and M83 $\Gamma$ results in Figure \ref{fig_gam_starburst}.  Interestingly, we find that while values of $\Gamma$ observed at lower \SigSFR\ ($<$0.1 \solperyr\ kpc$^{-2}$) are well-explained by $\Gamma$ relations with log(\tdep/yr) between $\sim$9--10, the starburst environments appear well-matched with a $\Gamma$ relation with log(\tdep/yr) of 8.0 (100 Myr).  This remarkable agreement between the predicted $\Gamma$ relation and the observations for starburst environments would be an intriguing success for the \citet{Kruijssen12} theoretical framework if measurements of \tdep\ in these systems prove to be consistent with the theoretically preferred value.

In the case of the M83 nuclear region, we can test the \tdep\ prediction of 100 Myr using published observations.  Based on \SigSFR\ and \SigGas\ measurements reported in \citet{Adamo15}, we derive a \tdep\ of $\sim$1 Gyr, which is a factor of 10 larger than the prediction.  This \tdep\ measurement would be worth revisiting as it is based on low spatial resolution CO observations \citep{Lundgren04} and a SFR derived from a \halpha\ luminosity.  In addition, variation of the CO-to-\Htwo\ conversion factor for the centers of galaxies could lead to an overestimation of \SigHtwo\ and \tdep\ \citep{Sandstrom13, Leroy13}.  Even considering these caveats, the likelihood of extremely short \tdep\ in these systems appears to be small.

If the observational \tdep\ constraints for these high \SigSFR\ environments are in fact longer than the 100 Myr value predicted in Figure \ref{fig_gam_starburst}, an alternative way to reproduce a steep $\Gamma$ relation at high \SigSFR\ is to include radiative pressure as an additional stellar feedback process.  As \citet{Kruijssen12} explored in their Appendix C, adding (or substituting) radiative feedback to the nominal supernova feedback prescription produces a $\Gamma$-\SigSFR\ relation with a different shape than obtained using supernova feedback alone --- particularly at high \SigSFR.  Using an alternative set of assumptions allowed by the \citet{Kruijssen12} code, we calculate a $\Gamma$ relation assuming combined feedback from supernova and radiative pressure (SN+Rad) and a characteristic value of log(\tdep/yr)=8.7 (500 Myr) and plot this relation for comparison in Figure \ref{fig_gam_starburst}.  The plot shows that this alternative theoretical solution also agrees well with the distribution of starburst $\Gamma$ observations.  The downside to this solution is that the relation predicted for SN+Rad feedback does not agree with $\Gamma$ and \tdep\ observations in non-starburst regions.  Therefore, some tuning of the model would be required, such that the contribution from radiative feedback would need to increase as a function of \SigSFR.

\begin{figure*}
\centering
\includegraphics[scale=0.7]{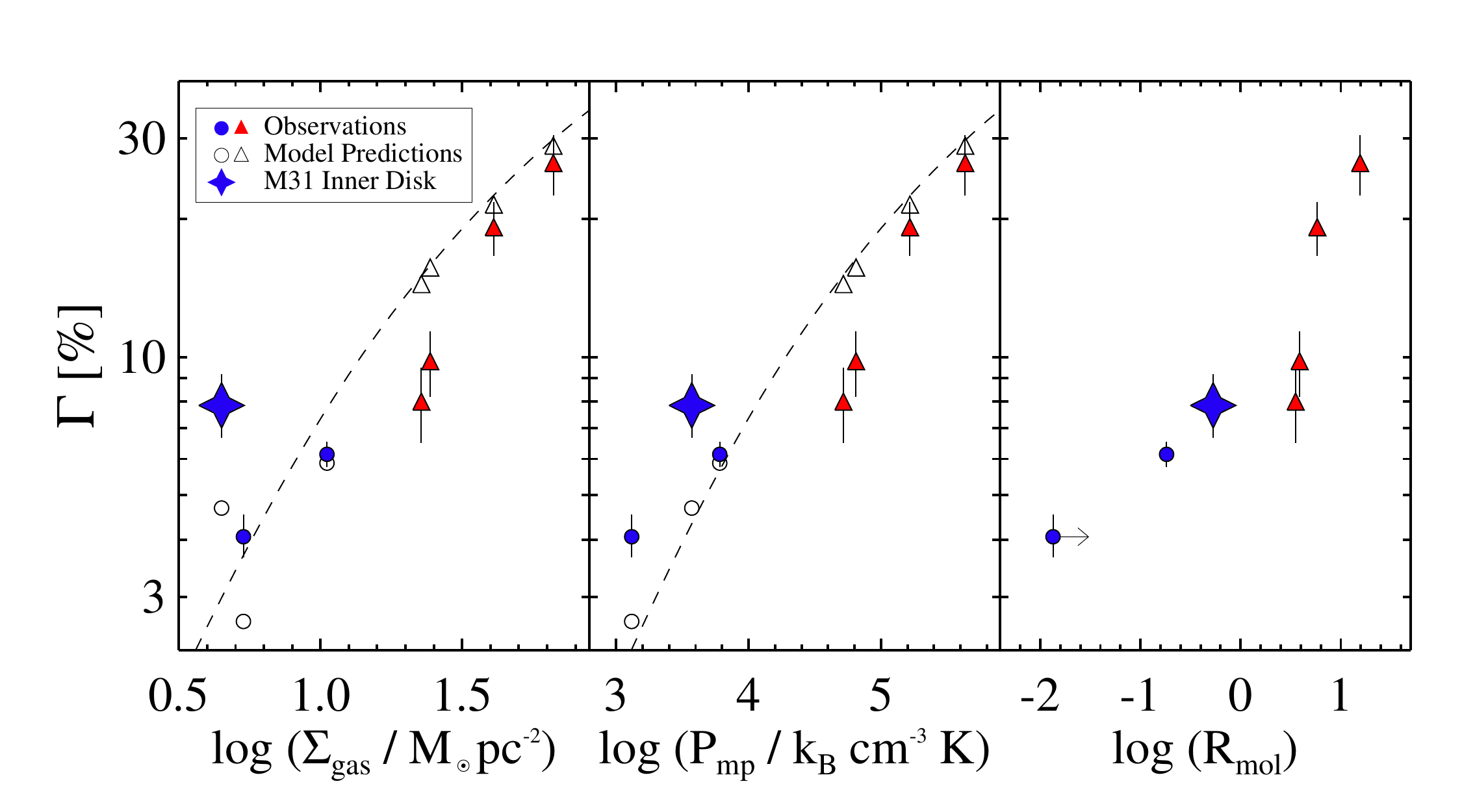}
\caption[$\Gamma$ Pressure Dependence]{Examination of M31 inner disk $\Gamma$ measurement (blue star) as a function of \SigGas\ (left), $P_{\rm mp}$ (center), and $R_{\rm mol}$ (\SigHtwo/\SigHI; right).  For comparison, we plot $\Gamma$ observations (filled symbols) and theoretical predictions (open symbols) for M31 (blue circles; this work) and M83 \citep[red triangles;]{Adamo15} analysis regions.  The left and center panels include the fiducial $\Gamma$ prediction curve (dashed line) from \citet{Kruijssen12} as well as individualized predictions derived using region-specific input parameters.  The data point for the M31 outer disk region in the right panel is represented by a lower limit on the molecular gas content due to the lack of short-spacing CO observations (see Section \ref{data_radio}).  We find better agreement between the M31 inner disk measurement and the underlying observed trend as we move from left to right in the figure.}
\label{fig_innerdisk}
\end{figure*}

Obtaining \tdep\ measurements for these high \SigSFR\ systems could help constrain models of cluster formation efficiency in starburst environments.  Fortunately these two proposed scenarios predict values of \tdep\ that differ by a factor of $\sim$5, which should produce an observationally detectable difference.  Additionally, further observations of $\Gamma$ behavior at high \SigSFR\ would provide a more complete picture of the variety and characteristics of long-lived cluster formation in starburst systems.

\subsection{Inner Disk $\Gamma$: Pressure Dependence} \label{discuss_innerdisk}

The high cluster formation efficiency obtained for the inner disk of M31 is notable.  We measure $\Gamma$ of $\sim$8\% for the region, which is unexpected considering it has the lowest gas surface density and the second lowest star formation rate surface density of all M31 regions.  In fact, the cluster formation efficiency of the inner disk is as large as those measured in the two most active star forming regions we studied in the 10 kpc ring (OB30/31 \& OB54; Regions 2a \& 2e).  In this section we explore possible explanations for the high $\Gamma$ observed in the inner disk, and use the unique attributes of the region to better understand the physical drivers that determine cluster formation efficiency.

We begin our investigation by comparing $\Gamma$ observations and theoretical predictions as a function of \SigGas\ in the left panel of Figure \ref{fig_innerdisk}.  We include M31 measurements as well as the M83 measurements from \citet{Adamo15}, ensuring our exploration covers the widest possible range of physical conditions.  For the M83 regions, we use $\Gamma$ and \SigGas\ measurements derived in Section \ref{discuss_gamres},
$\Omega$ constraints derived from \Htwo\ kinematics \citep{Lundgren04_kin}, \SigStar\ measurements \citep{Querejeta15}, stellar scale height constraints \citep{Herrmann09}, and assume a nominal 8 km s$^{-1}$ gas velocity dispersion to calculate $\Gamma$ predictions equivalent to those described in Section \ref{results_theory} (based on \SigGas, $\Omega$, $Q$, and $\phi_{P}$).

We plot $\Gamma$ observations and predictions for the combined set of M31 and M83 analysis regions in the left panel of Figure \ref{fig_innerdisk}, along with the fiducial prediction curve from \citet{Kruijssen12}.  The fact that $\Gamma$ observations and predictions do not increase monotonically with \SigGas\ demonstrates that while \SigGas\ is the primary input parameter driving the behavior of the \citet{Kruijssen12} model, it is not the physical parameter that best correlates with $\Gamma$.

We demonstrate in the center panel of Figure \ref{fig_innerdisk} that mid-plane pressure, rather than \SigGas, correlates most closely with $\Gamma$ in the \citet{Kruijssen12} model.  To create the plot, we estimate mid-plane pressure values for each region according to our Equation \ref{eq_pressure}, derived in \citet{KrumholzMcKee05}.  Focusing on the region-specific model predictions, we observe that the scatter about the fiducial relation seen among the M31 predictions in the \SigGas\ plot disappears in the \Pmp\ plot.  As discussed in Section \ref{results_theory}, the scatter in question traces back to variations in $\phi_{P}$, so the behavior of the theoretical predictions is not unexpected.  Nevertheless, the tight correlation of model predictions in the center panel of Figure \ref{fig_innerdisk} demonstrates that mid-plane pressure plays a primary role in setting the cluster formation efficiency in the \citet{Kruijssen12} model.  

In contrast to the theoretical predictions, the M31 and M83 $\Gamma$ observations do not yield a similarly tight correlation with mid-plane pressure.  The $\Gamma$ value for the inner disk of M31 still appears high relative to its associated \Pmp\ estimate, which falls between values for the outer disk and 10 kpc ring.  In addition, two of the M83 data points continue to fall below the predicted relation.

In the right most panel of Figure \ref{fig_innerdisk}, however, we find that the M31 inner disk region falls onto a monotonic relation between $\Gamma$ and the molecular fraction, $R_{\rm mol}$.  This result is somewhat surprising due to the fact that studies have shown that \Rmol\ and \Pmp\ are strongly correlated \citep[e.g., ][]{Blitz06, Leroy08}, so we expect similar $\Gamma$ behavior with respect to the two quantities.  The M31 inner disk's observed $R_{\rm mol}$ of $\sim$0.5 is typically associated with log (\Pmp\ / $k_{\rm B}$ cm$^{-3}$ K) $\sim$ 4.0$\pm$0.3 according to the empirical $R_{\rm mol}$--$P_{\rm mp}$ correlation published by \citet{Leroy08}.

We hypothesize that high mid-plane pressure in the inner disk of M31 could explain the relatively large observed values of $\Gamma$ and $R_{\rm mol}$ in the inner disk, despite the moderate initial estimate obtained for the region.  Assuming the \citet{Kruijssen12} model relation between \Pmp\ and $\Gamma$ shown in center panel of Figure \ref{fig_innerdisk}, it is possible to work backwards from the inner disk's $\sim$8\% $\Gamma$ measurement and obtain an estimate of log (\Pmp\ / $k_{\rm B}$ cm$^{-3}$ K) $\sim$ 4.  This prediction is a factor of 2.5 (0.4 dex) larger than the region's inferred mid-plane pressure, calculated using \SigGas\ estimates derived in Section \ref{data_radio} and the azimuthally-averaged stellar mass profile from \citet{Tamm12}, and agrees with the \Pmp\ value derived from the region's $R_{\rm mol}$.  

The large values of $\Gamma$ and $R_{\rm mol}$ observed in the inner disk of M31 are plausibly explained by a mid-plane pressure of $\sim$10$^4$ $k_{\rm B}$ cm$^{-3}$ K.  We speculate that M31's stellar bar \citep{Athanassoula06} could be responsible for the proposed, yet unaccounted for, mid-plane pressure in the M31 inner disk analysis region.  From a practical standpoint, the azimuthally-averaged \SigStar\ estimate used in our original mid-plane pressure calculation likely underestimates bar-enhanced stellar surface density in the inner disk.  Furthermore, bars are known to affect the dynamics of both stars and gas, leading to orbital crowding and an increase in cloud-cloud collisions at bar ends \citep[see e.g.,][]{Renaud15}.  While beyond the scope of this current work, the plausibility of our bar-driven inner disk pressure hypothesis could be tested using existing datasets that constrain the stellar distribution and kinematics of M31's inner disk.

We conclude that mid-plane pressure appears to play an important role in setting the cluster formation efficiency.  In the context of the \citet{Kruijssen12} model framework, increased pressure will tend to shift the gas density PDF to larger values, increasing the amount of mass found in the cluster producing tail of the distribution, resulting in an increase in cluster formation efficiency.  While a pressure-driven explanation for the large observed $\Gamma$ value in the inner disk is still unconfirmed, the strong correlation we uncovered between \Pmp\ and $\Gamma$ is a robust and useful result of this investigation.

\subsection{The Influence of Cluster Dissolution on $\Gamma$ Measurements} \label{discuss_dissolution}

Throughout this work, we assume that cluster dissolution has no effect on the 10--300 Myr PHAT cluster population we study.  As a result, we make no corrections to the cluster mass besides a cluster catalog completeness correction, assuming that the cluster population we see today is essentially unchanged since formation.  Here we review the points of evidence presented in this work that support this assumption.

The PHAT cluster age distribution presented in Figure \ref{fig_agemass} and discussed in Section \ref{analysis_cluster} shows a notable increase with logarithmic age, consistent with a constant formation history and little or no dissolution.  In the case of significant cluster dissolution, such as the d$N$/d$M \propto t^{-1}$ model advocated for in \citet{Fall09} and \citet{Fall12}, we would expect a uniform logarithmic age distribution, which is not consistent with the M31 young cluster population.

The agreement between $\Gamma$ determinations obtained for adjacent age ranges of 10--100 Myr and 100--300 Myr also provides compelling evidence that cluster dissolution has little effect on the derived values of $\Gamma$.  In the case of significant cluster dissolution, we expect results for the older age bin to show smaller $\Gamma$.  As we show in Figure \ref{fig_gam_agecomp} and discuss in Section \ref{results_gamma}, we find good agreement between the two age bins.  In fact, we find a small bias such that the 100--300 Myr $\Gamma$ measurements are a factor of 1.3 larger than the 10--100 Myr results on average.

While we conclude that cluster dissolution operates on sufficiently long timescales, such that the $\Gamma$ results presented here are unaffected, we do not rule out significant cluster destruction occurring on longer timescales.  Using a longer age baseline, we explore cluster dissolution timescales through detailed modeling of the cluster age and mass distribution for M31 in a separate work (M. Fouesneau, in preparation).

\section{Summary and Future Directions} \label{summary}

We conclude this work with a summary of the major contributions of this study, followed by a brief discussion about the broader implications of $\Gamma$ constraints and future directions for observational and theoretical progress.  The results of our observational $\Gamma$ work in M31 are summarized here:

\begin{enumerate}
\item We combine high quality cluster and field star formation history constraints from the PHAT survey, include detailed cluster and stellar completeness information in our calculations, and utilize a probabilistic modeling approach to perform the most detailed analysis of cluster formation efficiency ($\Gamma$) to-date.
\item We make spatially resolved measurements of $\Gamma$ across the disk of M31 and find values that vary between 4--8\%.  Our study significantly extends the range of environments for which observations of long-lived cluster formation efficiency have been obtained.
\item We apply knowledge about how the star formation relation behaves on sub-galactic scales, and differs between \Htwo\ to \HI-dominated star forming environments, and derive new predictions for spatially resolved $\Gamma$ observations as a function of \SigSFR.  The new $\Gamma$ relation flattens at low \SigSFR, in agreement with observations.
\item We derive new \tdep-dependent fiducial $\Gamma$ predictions to model $\Gamma$ observations in starburst environments. We propose an observational test to determine whether the theoretical $\Gamma$ model predictions using a \tdep=100 Myr star formation relation hold for starburst systems, or whether it is necessary to incorporate radiative feedback into the model for these systems.
\item We find good agreement between $\Gamma$ observations and theoretical predictions from \citet{Kruijssen12}, and demonstrate that mid-plane pressure is an important driver of cluster formation efficiency.
\end{enumerate}

Measurements of the fraction of stellar mass that is formed in long-lived star clusters as a function of star forming environment provide useful constraints towards understanding star formation behavior.  Following the interpretation of \citet{Kruijssen12}, these star clusters trace the stellar populations that are formed in environments where total star formation efficiencies (integrated over the lifetime of a star forming region, as opposed to per free-fall time) are high enough to produce stellar structures that survive gas expulsion during the transition out of an initial embedded phase.  Particularly when these $\Gamma$ measurements are combined with a characterization of the natal ISM, these observations paint an interesting picture connecting stellar feedback processes, formation efficiencies, and characteristics of the resulting stellar products.

We have only scratched the surface when it comes to using clusters and the spatial structure of newly formed stars to constrain star formation physics.  As pointed out in the review by \citet{Krumholz14_review}, the theoretical model for $\Gamma$ from \citet{Kruijssen12} can only predict the overall percentage of stellar mass locked up in long-lived clusters; it currently lacks the sophistication necessary to predict the mass function of these emergent clusters.  Work by \citet{Hopkins13} makes headway in predicting the spatial clustering of stars, therefore making predictions for the shape of the cluster mass function, but it also falls short of a complete treatment of cluster formation that accounts for the influence of stellar feedback on cluster outcomes.  In concurrence with \citet{Krumholz14_review}, we conclude that a theoretical understanding of long-lived cluster formation would benefit from the combination of theories that not only predict the overall fraction of bound mass, but also the distribution of that mass into the discrete systems we observe.  The cluster formation efficiency results presented here for M31, combined with mass function results presented in L. C. Johnson et~al.\ (2016, in preparation), will provide the most robust test of any such theory.

We are only beginning to utilize the full potential of $\Gamma$-based star formation studies.  As we discussed in Section \ref{discuss_gamres}, follow-up observations to characterize the star forming ISM in starburst systems would allow the differentiation between feedback mechanisms within the \citet{Kruijssen12} $\Gamma$ model framework.  Also, the growing number of well-constrained $\Gamma$ results span a wide variety of star forming environments and cluster formation activity.  However, as we saw in our study, sometimes it is exceptional regions like the inner disk of M31 that contribute significantly toward testing theoretical models.  Clearly, there is plenty of rewarding observational work still to be done.


\acknowledgements
{We acknowledge and thank the $\sim$30,000 Andromeda Project volunteers who made this research possible.  Their contributions are acknowledged individually at \url{http://www.andromedaproject.org/\#!/authors}.  We thank Nate Bastian, Luciana Bianchi, Yumi Choi, Dimitrios Gouliermis, Diederik Kruijssen, Tom Quinn, and the anonymous referee for their comments on the paper.  We are grateful to Robert Braun for providing us the \HI\ dataset.  Support for this work was provided by NASA through grant number HST-GO-12055 from the Space Telescope Science Institute, which is operated by AURA, Inc., under NASA contract NAS5-26555.  DRW is supported by NASA through Hubble Fellowship grant HST-HF-51331.01 awarded by the Space Telescope Science Institute.  This work made extensive use of NASAÕs Astrophysics Data System bibliographic services, as well as TOPCAT\footnote{\url{http://www.star.bris.ac.uk/~mbt/topcat/}} data visualization software.}

{\it Facilities:} \facility{HST (ACS, WFC3)}.


\appendix

\section{Catalog of Star Cluster Ages and Masses} \label{appendix_cat}

The full catalog of PHAT cluster fitting results will appear in A. Seth et~al.\ (in preparation).  In advance of this forthcoming publication, we present here the catalog of age and mass determinations for the 1249 star clusters utilized as part of this study.  Table \ref{tbl_cat} includes the cluster's Andromeda Project identifier \citep[referencing][]{Johnson15_AP}, age, mass, and analysis region membership.

\section{Calculating Average Surface Densities} \label{appendix_sigma}

Previous $\Gamma$ studies typically adopted a simple approach for deriving SFR and gas surface densities (\SigSFR\ and \SigGas) that used a single galaxy-wide aperture and measured area-averaged quantities.  There are a number of weaknesses in this approach.  First, the subjective definition of an outer boundary directly affects derived surface density values.  Adopting uniform definitions and procedures can serve to reduce these biases and uncertainties \citep[e.g., see discussion in Section 3.1.2 in][]{Adamo11}, but defining a outer limit for an inherently continuous distribution is difficult.  Second, area-averaged quantities assume a uniform intrinsic distribution, whereas star formation is inherently clumpy and irregular forming structures such as bars, arms, and rings.

The excellent spatial resolution available for all relevant M31 datasets allows us to compute surface densities using a deprojected 0.5 kpc$^2$ measurement kernel (with deprojected radius of $\sim$100 arcsec).  We calculate SFR-weighted average \SigSFR\ values to account for filling factor variations in the gas and SFR distributions, and explore how these stellar mass weighted values compare to the area-weighted metrics used in previous $\Gamma$ studies.

\begin{figure}[h]
\centering
\includegraphics[scale=0.6]{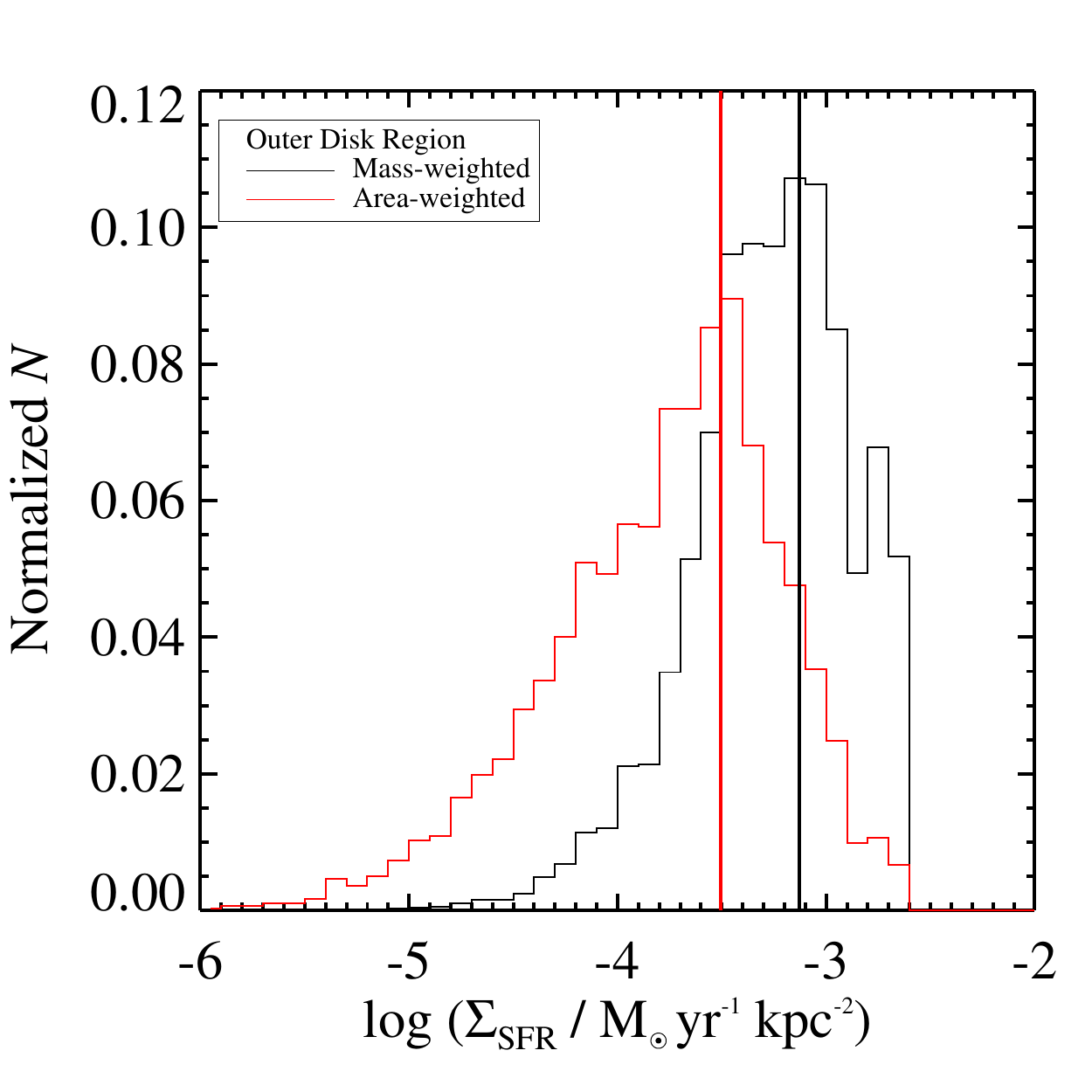}
\includegraphics[scale=0.6]{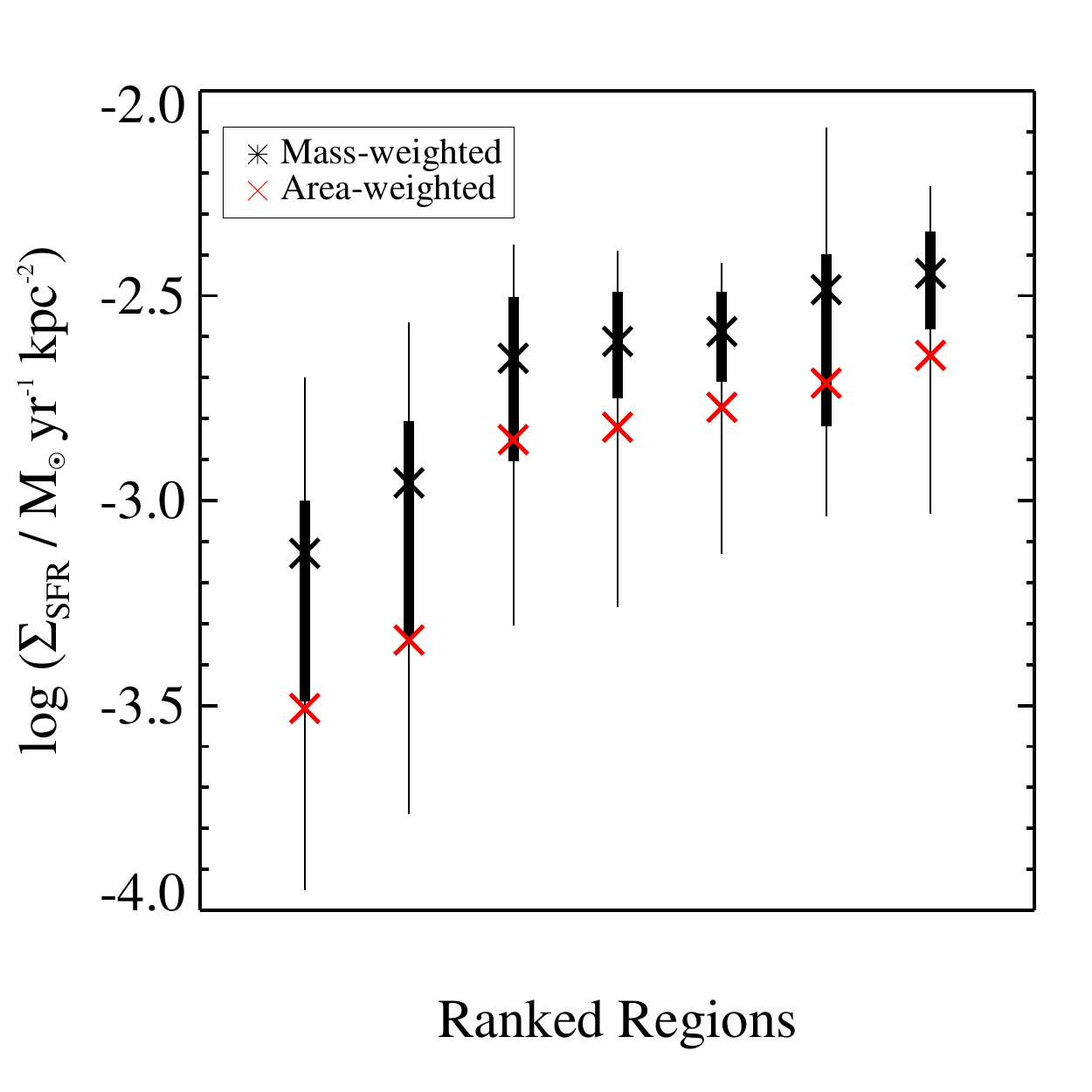}
\caption[Inter-region \SigSFR\ Distributions]{Left: We compare the \SigSFR\ distributions derived with SFR-weighting (black histogram) and without (area-weighted; red histogram) for the outer disk analysis region.  Thick vertical lines denote weighted (black) and unweighted (red) mean \SigSFR\ values.  Right: For each analysis region (ordered according to increasing \SigSFR) we plot the SFR-weighted mean \SigSFR\ values (black X), associated 25th--75th percentile range (thick black line), and 5th--95th percentile range (thin black line).  We compare these distributions to the area-weighted mean values (red X), showing the systematic difference between these estimates.}
\label{fig_sigsfrdist}
\end{figure}

We use the M31 outer disk (Region 2) to illustrate the difference between SFR and area-weighted \SigSFR\ measurements.  In the left panel of Figure \ref{fig_sigsfrdist}, we compare the distribution of 0.5 kpc$^2$ smoothed, unweighted (thus, area-weighted) \SigSFR\ measurements with the distribution of SFR-weighted measurements.  The thick vertical lines denote the area-weighted and SFR-weighted mean values; we report SFR-weighted mean values as our primary \SigSFR\ metric.  While the $\sim$0.4 dex relative difference in \SigSFR\ for the outer disk region is the biggest weighting-dependent difference among the seven M31 analysis regions (due to the relatively high contrasts between ring/arm and interarm/outskirts environments), a similar offset exists for all regions; we visualize these offsets in the right panel of Figure \ref{fig_sigsfrdist}.

It is also important to acknowledge that each of the M31 analysis regions contains a range of \SigSFR\ values.  We compute the interquartile range (from the 25th to 75th weighted percentile; thick line segments in right panel of Figure \ref{fig_sigsfrdist}) of the \SigSFR\ distribution for each analysis region, finding values from 0.2--0.4 dex.  While this is an unsurprising consequence of the clumpy, varying nature of star formation, it is important to keep in mind that characteristic mean \SigSFR\ values represent differences between broad underlying distributions of star formation intensities.

In addition to calculating \SigSFR, we also use a mass-weighted methodology to calculate robust measurements of \SigHI, \SigHtwo, and \SigGas.  M31's gas phase is dominated by a neutral \HI\ component, which has shown to have a low sub-kpc to kpc clumping factor \citep{Leroy13_clump}.  Therefore, measuring \SigGas\ using a 0.5 kpc$^2$ kernel provides accurate characterizations of intrinsic, \HI-dominated total gas densities.  This is not the case, however, when considering molecular gas and \SigHtwo\ alone.  High-resolution (20 pc) molecular gas observations obtained using CARMA (A. Schruba, in preparation) reveal molecular gas structures on the scale of 10--100 pc.  Therefore, one should take care when interpreting \SigHtwo\ values calculated in this work, as these densities are likely to be significantly diluted.


\section{$\Gamma$ Results from the Literature} \label{appendix_gamlit}

As discussed in Section \ref{discuss_gamgal}, we assemble a compilation of $\Gamma$ results from the literature to place M31 results in a broad context.  We present these literature results in Table \ref{tbl_lit}.  We are not the first to pursue this task \citep[e.g., see the recent compilation in Appendix B of][]{Adamo15}, however it is important to make clear the choices we have made in assembling (and in some cases, transforming) this set of results.  In particular, we make an effort to highlight where our choices differ from others.

We sought to utilize the highest quality results for individual galaxies when compiling this dataset.  In the case of M83 (NGC5236), we prefer the recent results from \citet{Adamo15} due to its near-complete coverage of the galaxy, surpassing previous constraints from \citet{SilvaVilla11} and \citet{SilvaVilla13}.

For the LMC, we utilize the recent $\Gamma$ result from \citet{Baumgardt13} of 15\%. This work makes use of updated cluster constraints from a compilation of sources for clusters with log(Mass/\solmass) $>$ 3.7.  We also update the far-IR luminosity-based integrated SFR estimate from \citet{Larsen00} with a CMD-based total star formation constraints from \citet{Harris09}.  Inferred masses in clusters and total stars and the derived $\Gamma$ increased significantly with respect to \citet{Goddard10} ($\Gamma = 5 \pm 0.5$\%), but agrees with the 10--20\% derived by \citet{Maschberger11} who use the same recent star formation history constraints.  We note that \citet{Baumgardt13} assume a power law with an index of -2.3 versus the traditional -2 for their cluster mass function extrapolation down to 100 \solmass.  An extrapolation using an index of -2 would give a result that was a factor of $\sim$0.7 smaller.  Also note that \citet{Baumgardt13} provides no accompanying \SigSFR\ value; we adopt the area normalization (79 kpc$^2$) used previously by \citet{Goddard10} to normalize the SFR (0.29 \solperyr).

In contrast to eliminating duplicate $\Gamma$ observations made on a common galaxy-wide scale, spatially resolved $\Gamma$ determinations provide unique constraints we do not want to ignore.  We tabulate individual spatially resolved $\Gamma$ constraints of the nuclear region of M83 from \citet{Goddard10}, as well as separate disk and nuclear measurements of NGC2997 from \citet{Ryon14}, but these constraints do not appear in Figure \ref{fig_gam_gal}, naturally, due to their sub-galaxy scale.  However, we omit these results from the presentation of spatially resolved results in Figure \ref{fig_gam_reg} due to a lack of available ISM constraints in the case of \citet{Ryon14}, and due to the $<$10 Myr age limitation of the \citet{Goddard10} result.

From the \citet{Cook12} dwarf galaxy work, we opt to use their ``binned'' 4--100 Myr $\Gamma$ results.  For the two age ranges they consider (4--10 Myr and 4--100 Myr), the authors combine the set of observed galaxies with $-4.5 <$ log \SigSFR\ $< -2.0$ into a single meta-galaxy.  This calculation serves to alleviate the problem of small numbers of clusters per individual galaxy (leading to large $\Gamma$ uncertainties), and to fold in galaxies that independently can only provide upper limit constraint on $\Gamma$.

We note that $\Gamma$ result for NGC4449 from \cite{Annibali11}, quoted for ages $<$10 Myr, depends completely on the inclusion or exclusion of the massive nuclear super star cluster; this single system hosts $>$70\% of the cluster mass considered in the $\Gamma$ calculation.  Similar to the behavior seen in the \citet{Cook12} results, this galaxy further demonstrates that the stochastic nature of star formation in dwarf galaxies can lead to large variations in the derived result.  Further, the $\Gamma$ calculation in this work uses a mass function extrapolation assuming a power law form with -2 slope, down to a lower mass limit of 1000 \solmass.  A correction factor of $\sim$1.4 could be applied to bring the data in line with the standard 100 \solmass\ assumption, but we opt to tabulate and plot the work's original values.

We utilize the ``P1'' results from \citet{SilvaVilla10, SilvaVilla11} that do not include dissolution modeling, as opposed to their mass independent destruction (MID) or mass dependent destruction (MDD) constraints.  These results, which were also used by \citet{Cook12}, provides a better match to the model-independent, empirical approach of the other studies with which we compare.  In addition, we note that these results were calculated using a mass function extrapolation assuming a Schechter function with $m_c=2\times10^5$ \solmass\ down to a lower mass limit of 10 \solmass, which differs from the canonical value of 100 \solmass\ used in other $\Gamma$ studies.  Similar to the case of NGC4449 discussed previously, a correction factor of $\sim$0.8 could be applied to bring the data in line with the standard minimum cluster mass assumption, but we opt to tabulate and plot the work's original values.  In addition, this work utilizes an age-dependent observational completeness limit for mass function extrapolation, and includes a scaling factor applied to the observed CFR to account for coverage differences between data used for cluster fitting versus that used for total star formation fitting of the field populations.

Finally, we would like to highlight two cases where galaxy-integrated $\Gamma$ constraints deviate strongly from the observed: IC10 and NGC45.  Both of these galaxies are dwarf systems with relatively low integrated SFRs.  These low SFRs and small number statistics among the observed clusters imply large $\Gamma$ uncertainties due to stochastic sampling of the cluster mass function.  Unfortunately, neither of these results were accompanied by reported uncertainties \citep[although, uncertainty estimates for NGC45 were made available for MID and MDD based results by][]{SilvaVilla11}.  We also wish to highlight that \citet{SilvaVilla13} mentions the possibility that a number of ancient massive globular cluster systems were assigned integrated light-based ages that erroneously placed them in the 10--100 Myr range used to determine $\Gamma$.  This case serves as an example that, particularly in the case of small numbers of clusters, errors stemming from a variety of sources (many of which are not accounted for in uncertainty calculations) can contribute to the large scatter in reported $\Gamma$ results.

This literature sample provides $\Gamma$ constraints from 30 galaxies, combining measurements from 19 individual galaxies and 11 dwarf galaxies that are analyzed together by \citet{Cook12}.  Note that this total does not count the upper limits contributed by 23 additional dwarf galaxies in \citet{Cook12} that have no young clusters detected.


\section{A Star Formation Relation Based on Spatially Resolved Observations} \label{appendix_sflaw}

Here we report the detailed specifications for the star formation relation defined in Section \ref{discuss_gamres} based on spatially resolved observations from \citet{Bigiel08}.  We use three two-component power law functions to define a median relation and accompanying upper and lower envelope relations, as plotted in Figure \ref{fig_ks}.  The median relation is defined to agree with star formation relation (\SigSFR\ $\propto$ \SigGas$^N$) results from \citet{Leroy13} in the molecular-dominated high density regime ($N=1.0$), and track the transition to \HI-dominated star formation environments using a steeper slope ($N=3.3$) for \SigGas\ $\lesssim$ 10 \solperpc.  We define this median relation as:
\begin{equation}
\frac{\Sigma_{\rm SFR}}{M_{\sun} \text{ yr}^{-1} \text{ kpc}^{-2}} = 
\begin{cases}
	1\times 10^{-3.4} \left(\frac{\Sigma_{\rm gas}}{M_{\sun} \rm{pc}^{-2}} \right), & \text{if } 1.0 \leq \log \left(\frac{\Sigma_{\rm gas}}{M_{\sun} \rm{pc}^{-2}} \right) < 2.0 \\
	1\times 10^{-5.93} \left(\frac{\Sigma_{\rm gas}}{M_{\sun} \rm{pc}^{-2}} \right)^{3.3}, & \text{if } 0.3 \le \log \left(\frac{\Sigma_{\rm gas}}{M_{\sun} \rm{pc}^{-2}} \right) \le 1.0.
\end{cases}
\end{equation}
We bracket the median relation with an upper and lower envelope that are chosen to reproduce the spread in \tdep\ observed by \citet{Bigiel08}.  We define the upper envelope as:
\begin{equation}
\frac{\Sigma_{\rm SFR}}{M_{\sun} \text{ yr}^{-1} \text{ kpc}^{-2}} = 
\begin{cases}
	1\times 10^{-3.1} \left(\frac{\Sigma_{\rm gas}}{M_{\sun} \rm{pc}^{-2}} \right), & \text{if } 1.0 \leq \log \left(\frac{\Sigma_{\rm gas}}{M_{\sun} \rm{pc}^{-2}} \right) < 2.0\\
	1\times 10^{-5.055} \left(\frac{\Sigma_{\rm gas}}{M_{\sun} \rm{pc}^{-2}} \right)^{3.3}, & \text{if } 0.3 \le \log \left(\frac{\Sigma_{\rm gas}}{M_{\sun} \rm{pc}^{-2}} \right) \le 1.0.
\end{cases}
\end{equation}
The lower envelope is defined as:
\begin{equation}
\frac{\Sigma_{\rm SFR}}{M_{\sun} \text{ yr}^{-1} \text{ kpc}^{-2}} = 
\begin{cases}
	1\times 10^{-3.7} \left(\frac{\Sigma_{\rm gas}}{M_{\sun} \rm{pc}^{-2}} \right), & \text{if } 1.0 \leq \log \left(\frac{\Sigma_{\rm gas}}{M_{\sun} \rm{pc}^{-2}} \right) < 2.0\\
	1\times 10^{-6.69} \left(\frac{\Sigma_{\rm gas}}{M_{\sun} \rm{pc}^{-2}} \right)^{3.3}, & \text{if } 0.3 \le \log \left(\frac{\Sigma_{\rm gas}}{M_{\sun} \rm{pc}^{-2}} \right) \le 1.0.
\end{cases}
\end{equation}
The median, upper envelope, and lower envelope relations have inflection points (log \SigGas, log \SigSFR) at (1.1, -2.3), (0.85, -2.25), and (1.3, -2.4), respectively.  We define these relations over the range of \SigGas\ parameter space spanned by the observations: $0.3 \le \log (\Sigma_{\rm gas}/M_{\sun} \rm{pc}^{-2}) \le 2.0$.  Outside this range of total gas densities, observations tentatively point to qualitatively different behavior.  For gas densities $<$2 \solperpc, \citet{Bigiel10} presents evidence for a flattening of the star formation relation that hints at an asymptotic \tdep\ value of $\sim$10$^{11}$ yr.  For gas densities $>$100 \solperpc, a starburst mode of star formation likely prevails \citep[e.g., see][]{Daddi10}, characterized by \tdep\ on the order of $\sim$10$^{7}$--10$^{8}$ yr and a slope of $N$$\sim$1.3--1.4.  Therefore, extrapolation of this relation beyond the adopted \SigGas\ limits is not advised.


\bibliographystyle{apj}
\bibliography{../clusterlit}



\begin{deluxetable*}{lcccccccccc}
\tabletypesize{\footnotesize}
\tablecaption{ISM Observational Data \label{tbl_ismobs}}

\tablehead{
\colhead{Region} & \colhead{Region} & \colhead{\SigHI} & \colhead{\SigHtwo} & \colhead{\SigGas} & \colhead{$R_{\rm mol}$} & \colhead{$\sigma_{\rm gas}$} & \colhead{$\overline{R}_{\rm gc}$\tablenotemark{a}} & \colhead{$\Omega$} & \colhead{$Q$} & \colhead{$\phi_{P}$} \\
\colhead{ID} & \colhead{Name} & \colhead{(\solperpc)} & \colhead{(\solperpc)} & \colhead{(\solperpc)} & \colhead{} & \colhead{(km s$^{-1}$)} & \colhead{(kpc)} & \colhead{(Myr$^{-1}$)} & \colhead{} & \colhead{}
}

\startdata
1  & Inner Disk      & 2.44 & 2.03 &  4.47 & 0.532 & 7.94 &  6.61 & 0.031 & 5.65 & 5.6 \\
2  & Ring-Total      & 8.38 & 2.16 & 10.54 & 0.182 & 8.65 & 11.80 & 0.021 & 1.77 & 1.6 \\
2a & Ring-OB30/31    & 7.82 & 2.21 & 10.02 & 0.218 & 9.73 & 11.45 & 0.022 & 2.19 & 1.7 \\
2b & Ring-OB39/40/41 & 9.40 & 2.61 & 12.01 & 0.208 & 9.46 & 12.14 & 0.021 & 1.70 & 1.6 \\
2c & Ring-OB48       & 9.42 & 1.92 & 11.34 & 0.145 & 8.03 & 12.11 & 0.021 & 1.53 & 1.6 \\
2d & Ring-Spur       & 8.17 & 1.66 &  9.84 & 0.123 & 8.16 & 12.16 & 0.021 & 1.79 & 1.7 \\
2e & Ring-OB54       & 7.37 & 2.29 &  9.66 & 0.259 & 8.12 & 11.14 & 0.023 & 1.98 & 1.7 \\
3  & Outer Disk      & 5.17 & 0.18 &  5.34 & 0.013 & 7.17 & 15.83 & 0.016 & 2.20 & 1.4 \\
\nodata  & Total           & 6.08 & 2.08 &  8.17 & 0.186 & 8.12 & \nodata & 0.021 & 2.14 & 1.8
\enddata

\tablenotetext{a}{Mass-weighted mean galactocentric radius.}

\end{deluxetable*}


\begin{deluxetable*}{lcccccc}
\tabletypesize{\footnotesize}
\tablecaption{Cluster and SFH Observational Data \label{tbl_obs}}

\tablehead{
\colhead{Region} & \colhead{Region} & \colhead{$M_{\rm cl, obs}$} & \colhead{$M_{\rm tot}$} & \colhead{log \SigSFR} & \colhead{$m_{\rm lim}$\tablenotemark{a}} & \colhead{$s_{\rm lim}$\tablenotemark{b}} \\
\colhead{ID} & \colhead{Name} & \colhead{(10$^4$ \solmass)} & \colhead{(10$^6$ \solmass)} & \colhead{(\solperyr\ kpc$^{-2}$)} & \colhead{(\solmass)} & \colhead{}
}

\startdata
\multicolumn{7}{c}{10--100 Myr} \\
\hline
1  & Inner Disk      &  11.36 $\pm$ 0.35 &   3.58 $\pm$ 0.04 & -2.96 &  946 & 3.5 \\
2  & Ring-Total      &  52.93 $\pm$ 0.50 &  19.80 $\pm$ 0.16 & -2.55 &  741 & 4.2 \\
2a & Ring-OB30/31     &  16.77 $\pm$ 0.30 &   4.62 $\pm$ 0.09  & -2.45 &  687 & 4.8 \\
2b & Ring-OB39/40/41     &   7.45 $\pm$ 0.19 &   3.49 $\pm$ 0.07 & -2.59 &  749 & 4.1 \\
2c & Ring-OB48     &   7.16 $\pm$ 0.22 &   3.12 $\pm$ 0.06 & -2.61 &  697 & 4.4 \\
2d & Ring-Spur       &   8.62 $\pm$ 0.19 &   4.79 $\pm$ 0.07 & -2.65 &  721 & 4.5 \\
2e & Ring-OB54        &  12.92 $\pm$ 0.21 &   3.77 $\pm$ 0.08 & -2.48 &  830 & 7.0 \\
3  & Outer Disk      &   6.65 $\pm$ 0.18 &   3.49 $\pm$ 0.04 & -3.13 &  522 & 6.3 \\
\nodata  & Total           &  70.93 $\pm$ 0.64 &  26.87 $\pm$ 0.16 & -2.63 &  740 & 5.0 \\
\hline
\multicolumn{7}{c}{100--300 Myr} \\
\hline
1  & Inner Disk      &  34.38 $\pm$ 0.68 &   8.77 $\pm$ 0.14 & -3.03 &  953 &  6.3 \\
2  & Ring-Total      & 104.10 $\pm$ 3.57 &  45.23 $\pm$ 0.52 & -2.53 & 1130 &  6.5 \\
2a & Ring-OB30/31    &  12.05 $\pm$ 3.43 &  10.29 $\pm$ 0.28 & -2.42 & 1226 &  7.5 \\
2b & Ring-OB39/40/41 &  16.33 $\pm$ 0.31 &   7.90 $\pm$ 0.21 & -2.60 & 1249 &  5.7 \\
2c & Ring-OB48       &  16.47 $\pm$ 0.31 &   7.26 $\pm$ 0.18 & -2.61 & 1145 &  7.1 \\
2d & Ring-Spur       &  34.50 $\pm$ 0.66 &  10.72 $\pm$ 0.23 & -2.62 &  805 &  6.1 \\
2e & Ring-OB54       &  24.73 $\pm$ 0.55 &   9.06 $\pm$ 0.23 & -2.46 & 1146 &  6.0 \\
3  & Outer Disk      &  18.96 $\pm$ 0.64 &   7.36 $\pm$ 0.10 & -3.29 &  651 &  6.0 \\
\nodata  & Total           & 157.40 $\pm$ 3.69 &  61.36 $\pm$ 0.54 & -2.62 & 1086 &  6.2
\enddata

\tablenotetext{a}{The 50\% cluster catalog mass completeness limit.}
\tablenotetext{b}{The logistic slope parameter for completeness function.}
\end{deluxetable*}


\begin{deluxetable*}{lcccc}
\tabletypesize{\footnotesize}
\tablecaption{$\Gamma$ Results and Predictions \label{tbl_gamma}}

\tablehead{
\colhead{Region} & \colhead{Region} & \colhead{$\Gamma_{\rm 10-100}$} & \colhead{$\Gamma_{\rm 100-300}$} & \colhead{$\Gamma_{\rm predict}$} \\
\colhead{ID} & \colhead{Name} & \colhead{(\%)} & \colhead{(\%)} & \colhead{(\%)}
}

\startdata
1  & Inner Disk      & $7.9^{+1.3}_{-1.2}$ & $9.2^{+0.8}_{-0.8}$ & 4.7 \\
2  & Ring-Total      & $6.1^{+0.4}_{-0.4}$ & $6.4^{+0.5}_{-0.4}$ & 5.9 \\
2a & Ring-OB30/31    & $7.6^{+0.7}_{-0.7}$ & $3.6^{+0.7}_{-0.6}$ & 5.8 \\
2b & Ring-OB39/40/41 & $5.1^{+0.8}_{-0.6}$ & $6.5^{+1.0}_{-0.9}$ & 6.6 \\
2c & Ring-OB48       & $5.4^{+0.8}_{-0.7}$ & $6.5^{+0.9}_{-0.8}$ & 6.1 \\
2d & Ring-Spur       & $4.4^{+0.6}_{-0.5}$ & $6.6^{+0.5}_{-0.5}$ & 5.5 \\
2e & Ring-OB54       & $8.0^{+0.9}_{-0.9}$ & $7.3^{+0.9}_{-0.8}$ & 5.4 \\
3  & Outer Disk      & $4.1^{+0.5}_{-0.4}$ & $5.3^{+0.5}_{-0.4}$ & 2.7 \\
\nodata  & Total           & $5.9^{+0.3}_{-0.3}$ & $6.6^{+0.4}_{-0.3}$ & 4.8
\enddata

\end{deluxetable*}


\begin{deluxetable*}{lccccccc}
\tabletypesize{\footnotesize}
\tablecaption{Cluster Fitting Results \label{tbl_cat}}

\tablehead{
\colhead{AP ID} & \colhead{Region ID} & \colhead{} & \colhead{log (Age/yr)} & \colhead{} & \colhead{} & \colhead{log (Mass/\solmass)} & \colhead{} \\
\colhead{} & \colhead{} & \colhead{Best} & \colhead{P16} & \colhead{P84} & \colhead{Best} & \colhead{P16} & \colhead{P84}
}

\startdata
2 & 2c & 8.4 & 8.4 & 8.4 & 3.98 & 3.95 & 3.98 \\
5 & 1 & 8.4 & 8.3 & 8.4 & 3.41 & 3.40 & 3.45 \\
7 & 3 & 8.2 & 7.9 & 8.2 & 3.16 & 3.12 & 3.16 \\
14 & 2e & 8.2 & 8.2 & 8.3 & 4.08 & 4.08 & 4.13 \\
16 & 2a & 8.4 & 8.4 & 8.4 & 3.57 & 3.54 & 3.60
\enddata

\tablecomments{Table \ref{tbl_cat} is published in its entirety in the electronic edition of the {\it Astrophysical Journal}.  A portion is shown here for guidance regarding its form and content.}

\end{deluxetable*}


\begin{deluxetable*}{lccccc}
\tablecolumns{6}
\tabletypesize{\footnotesize}
\tablecaption{$\Gamma$ Results from the Literature \label{tbl_lit}}

\tablehead{
\colhead{Galaxy} & \colhead{\SigSFR} & \colhead{$\Gamma$} & \colhead{Reference} \\
\colhead{} & \colhead{(\solperyr\ kpc$^{-2}$)} & \colhead{(\%)} & \colhead{}
}

\startdata
\multicolumn{6}{c}{Galaxy Integrated Measurements} \\
\hline
  NGC1569 & 0.03 & 13.9 $\pm$ 0.8 & \citealt{Goddard10} \\
  NGC3256 & 0.62 & 22.9$^{+7.3}_{-9.8}$ & \citealt{Goddard10} \\
  NGC6946 & 0.0046 & 12.5$^{+1.8}_{-2.5}$ & \citealt{Goddard10} \\
  SMC & 0.001 & 4.2$^{+0.2}_{-0.3}$ & \citealt{Goddard10} \\
  Milky Way & 0.012 & 7.0$^{+7}_{-3.0}$ & \citealt{Goddard10} \\
  ESO338 & 1.55 & 50.0 $\pm$ 10.0 & \citealt{Adamo11} \\
  Haro 11 & 2.16 & 50.0$^{+13}_{-15}$ & \citealt{Adamo11} \\
  ESO185-IG13 & 0.52 & 26.0 $\pm$ 5.0 & \citealt{Adamo11} \\
  MRK930 & 0.59 & 25.0 $\pm$ 10.0 & \citealt{Adamo11} \\
  SBS0335-052E & 0.95 & 49.0 $\pm$ 15.0 & \citealt{Adamo11} \\
  NGC45 & 0.00101 & 17.3 & \citealt{SilvaVilla11} \\
  NGC1313 & 0.011 & 9.0 & \citealt{SilvaVilla11} \\
  NGC4395 & 0.00466 & 2.6 & \citealt{SilvaVilla11} \\
  NGC7793 & 0.00643 & 9.8 & \citealt{SilvaVilla11} \\
  NGC4449 & 0.04 & 9.0 & \citealt{Annibali11} \\
  ANGST Dwarfs ($<$100 Myr) & 3e-5--1e-2 & 1.65 & \citealt{Cook12} \\
  LMC & 0.00366 & 15.0 & \citealt{Baumgardt13} \\
  NGC2997 & 0.0094 & 10.0 $\pm$ 2.6 & \citealt{Ryon14} \\
  IC10 & 0.03 & 4.2 & \citealt{Lim15} \\
  M83 (0.45--4.5 kpc)\tablenotemark{a} & 0.019 & 12.5 $\pm$ 1.4 & \citealt{Adamo15} \\
\hline
\multicolumn{6}{c}{Spatially Resolved Measurements} \\
\hline
  M83 (Nuclear) & 0.54 & 26.7$^{+5.3}_{-4.0}$ & \citealt{Goddard10} \\
  NGC2997 (Disk) & 0.0049 & 7.0 $\pm$ 2.0 & \citealt{Ryon14} \\
  NGC2997 (Nuclear) & 0.164 & 12.0 $\pm$ 4.0 & \citealt{Ryon14} \\
  M83 (0.45--2.3 kpc)\tablenotemark{a} & 0.013 & 26.5 $\pm$ 4.0 & \citealt{Adamo15} \\
  M83 (2.3--3.2 kpc)\tablenotemark{a} & 0.028 & 19.2 $\pm$ 2.6 & \citealt{Adamo15} \\
  M83 (3.2--3.9 kpc)\tablenotemark{a} & 0.022 & 9.8 $\pm$ 1.6 & \citealt{Adamo15} \\
  M83 (3.9--4.5 kpc)\tablenotemark{a} & 0.014 & 8.0 $\pm$ 1.5 & \citealt{Adamo15}
\enddata

\tablenotetext{a}{We utilize the 10--50 Myr $\Gamma$ results from \citet{Adamo15}.}

\end{deluxetable*}


\end{document}